\documentclass[a4paper,fleqn,usenatbib]{mnras}

\usepackage{newtxtext,newtxmath}

\usepackage[T1]{fontenc}
\usepackage{ae,aecompl}

\usepackage{graphicx}	
\usepackage{amsmath}	
\usepackage{amssymb}	

\title[Tidal stripping to place MGC1 in M31]{Tidal stripping as a mechanism for placing globular clusters on wide orbits: the case of MGC1 in M31}

\author[E. P. Andersson et al.]{Eric P. Andersson,$^{1}$\thanks{E-mail: eric@astro.lu.se}
Melvyn B. Davies,$^{1}$
\\
$^{1}$Lund Observatory, Department of Astronomy and Theoretical Physics, Box 43, SE 221-00 Lund, Sweden \\
}

\date{Accepted XXX. Received YYY; in original form ZZZ}

\pubyear{2018}

\begin{document}
\label{firstpage}
\pagerange{\pageref{firstpage}--\pageref{lastpage}}
\maketitle

\begin{abstract}
The globular clusters of large spiral galaxies can be divided into two populations: one which formed in-situ and one which comprises clusters tidally stripped away from other galaxies. In this paper we investigate the contribution to the outer globular cluster population in the M31 galaxy through donation of clusters from dwarf galaxies. We test this numerically by comparing the contribution of globular clusters from simulated encounters to the observed M31 globular cluster population. To constrain our simulations, we specifically investigate the outermost globular cluster in the M31 system, MGC1. The remote location of MGC1 favours the idea of it being captured, however, the cluster is devoid of features associated with tidal interactions. Hence we separate simulations where tidal features are present and where they are hidden. We find that our simulated encounters can place clusters on MGC1-like orbits. In addition, we find that tidal stripping of clusters from dwarf galaxies leaves them on orbits having a range of separations, broadly matching those observed in M31. We find that the specific energies of globular clusters captured by M31 closely matches those of the incoming host dwarf galaxies. Furthermore, in our simulations we find an equal number of accreted clusters on co-rotating and counter-rotating orbits within M31 and use this to infer the fraction of clusters that has been accreted. We find that even close in roughly $50\%$ of the clusters are accreted, whilst this figure increases to over $80\%$ further out.
\end{abstract}

\begin{keywords}
globular clusters: general -- galaxies: individual: M31
\end{keywords}



\section{Introduction}
Globular clusters are believed to be an integral part in galaxy evolution, see e.g, \citet{Brodie2006, Renaud2018}, and understanding the origin of different globular clusters is essential to fully understanding their formation. There are two viable mechanisms that can populate a galaxy with clusters; either the globular clusters form {\it in-situ}, that is from gaseous material inside the galaxy itself, or they form outside and then are accreted to the total population. \citet{SearleZinn1978} suggested this twofold build-up as a hypothesis for explaining the bimodalities observed in the Milky Way cluster population \citep[see, e.g.,][]{Zinn1985,Harris1996,MarinFranch2009}, and many works has investigated this scenario since then \citep[see, e.g.,][]{Cote1998,Tonini2013,Renaud2017}. 

The Milky Way is not the only galaxy that shows such bimodalities, in fact bimodalities are very common in observed galaxies, see e.g. \citet{Brodie2006} for a review. Most often this is referred to as the blue and red clusters, due to the globular clusters showing a separation in colour \citep[see, e.g.,][]{GebhardtKissler-Patig1999,Larsen2001,KunduWhitmore2001,Peng2006}. Furthermore, in the Milky Way the blue clusters has been linked to a metal-poor population that is spatially extended without clear signs of rotating, whereas the red clusters are more metal-rich, spatially concentrated and shows signs of co-rotating with the galaxy, see \citet{ArmandroffZinn1988, Cote1999}. The spatial distribution and kinematics of the two populations are in favour of building the blue population by accretion, while the red population forms {\it in-situ}.

A more compelling evidence for the two-fold build up is the on-going accretion of globular clusters from satellite galaxies in the Universe today. One example is the Sagittarius dwarf, which is believed to have donated multiple globular clusters to the Milky Way in the past billion years, see \citet{Layden2000, Siegel2007}. Further evidence of accretion events can also be found in the outer parts of our neighbouring galaxy Andromeda (henceforth M31), which was well studied in the Pan-Andromeda-Archaeological-Survey (PAndAS), \citep[see, e.g.,][]{Martin_et.al2006,Ibata+2007,McConnachie2008}. Using the PAndAS data, \citet{Machey_Huxor2010} found that there is a striking alignment between the position of outer globular cluster and tidal debris, such as stellar streams. By detailed investigation of the likelihood of chance alignment, \citeauthor{Machey_Huxor2010} concluded that the outer globular cluster population consists of $\goa 80\%$ accreted clusters. 

In this work we focus on the globular cluster MGC1, which is a globular cluster on an extremely wide orbit in M31. \citet{Martin_et.al2006} discovered this particular cluster for which they determined a position on the sky with right ascension $0^{\rm h}50^{\rm m}42.5^{\rm s}$ and declination $+32\degr54\arcmin59.6\arcsec$. This gives MGC1 a projected distance of $\approx117\,{\rm kpc}$ from the M31 centre. In another work by \citet{MackeyFerguson2010}, the authors observed MGC1 using the Gemini/GMOS observatory to determine a precise distance to the cluster. They found a distance modulus of $\mu = 23.95\pm0.06$, which gives MGC1 a radial distance from M31 of $200\pm20\,{\rm kpc}$. This makes MGC1 the most remote cluster in the Local Group with some considerable margin. Although a proper motion has not yet been determined for MGC1, several authors have measured its radial velocity, see, e.g., \citet{Galleti2007,Alves-Brito2009}. Comparing to the systematic velocity of M31 \citep{vanDerMarel2012}, the MGC1 has a radial velocity that is comfortably lower than the M31 escape velocity \citep{Chapman2007}. Thus, MGC1 is likely one of M31 {\it bona fide} members and not an intergalactic cluster. 

The metallicity of the cluster has been somewhat debated, where the authors of the original paper, \citet{Martin_et.al2006} determined a metallicity $[{\rm Fe/H}]\approx-1.3$ by fitting isochrones to the its colour-magnitude diagram (CMD). Using spectra from the Keck/HIRES instrument, \citet{Alves-Brito2009} argued that the metallicity for MGC1 was $[{\rm Fe/H}]\approx-1.37\pm 0.15$, which is in agreement with \citeauthor{Martin_et.al2006}. Controversially, \citet{MackeyFerguson2010} derived a significantly lower metallicity at a reported value of $[{\rm Fe/H}]\approx-2.7$. Although, said authors do not agree on the exact value, they all coincide with a metallicity that would contribute to the blue, metal-poor population of the Milky Way. For the M31 such bimodality is not clear, where both cases has been argued \citep[see, e.g.,][]{Barmby+2000,Fan+2008,Caldwell+2011}. 

The remote location and low metallicity of MGC1 is in favour of an accretion origin for the cluster, however there are restrictions to this scenario as well. \citet{MackeyFerguson2010} investigated this possibility by searching for signs of tidal interaction. MGC1 does coincide spatially with three satellite galaxies, however, this is due to projection where the three satellites reside significantly closer to the M31. Additionally, the three galaxies and MGC1 shows different velocities and are therefore likely on very different orbits. Another tracer for recent tidal interaction is stellar streams around the object. \citet{MackeyFerguson2010} used star counts in the vicinity of MGC1 to look for such features but could not find any. Moreover, \citeauthor{MackeyFerguson2010} found that MGC1 has an extremely extended structure, with stars that match the MGC1 stellar population out to $450\,{\rm pc}$ and possibly even $900\,{\rm pc}$. This is a significant fraction of the tidal radius at the location of MGC1, thus it must have spent some considerable time in isolation. 

In this work we address the question of whether it is possible that MGC1 did in fact originate from an accretion event. We test this by using numerical simulations of encounters between dwarf galaxies and the M31, in which we look at the probability of tidally stripping away globular clusters from the dwarf galaxy and place them on wide enough orbits for them to be observed as the MGC1. Furthermore, we constrain our encounters as to not leave visible traces by using an analytic expression of the tidal radius to check whether the dwarf galaxy or the clusters sustain significant damage to its luminous matter during the encounter. 

The report is structured with a summary of the numerical scheme and brief review of the tidal radius in Section~\ref{sec:methodology}, followed by results and their implications in Section~\ref{sec:results}, first focusing on a typical encounter in our simulation and then looking at how the likelihood of obtaining a cluster similar to MGC1 depends on the dwarf galaxy properties and orbital parameters in all encounter that where simulated. In Section~\ref{sec:building_M31_GCpop} we investigate what our results imply for the observed globular cluster population in M31 by comparing the general properties of clusters captured by M31 on orbits other than those that resemble the orbital properties of MGC1 to those observed in the accreted cluster population of M31. In Section~\ref{sec:discussion}, we discuss our results, the limitations of our model and suggest possible histories of MGC1 other than that covered in this work for future work. Finally, in Section~\ref{sec:conclusion} we list the major findings of this work. 

\section{Methodology}\label{sec:methodology}
We use numerical integration to investigate whether globular clusters can be tidally stripped away from dwarf galaxies during encounters with M31, after which they should be left on wide orbits around M31, such that they can be observed as MGC1. Additionally, the encounters are restricted to not cause visible damage to the luminous structure of M31 or the dwarf galaxy due to the absence of tidal features around the observed MGC1. We use a Runge-Kutta integrator scheme with an adaptive time-step \citep{Cash1990} for which the acceleration is calculated from a static, axisymmetric, three component potential field for M31 and a moving, spherically symmetric potential, for the dwarf galaxy. These potential fields are described in Section~\ref{sec:potential_models}. Out of interest in simplicity the globular clusters were treated as test-particles in all simulations. For each encounter we place a thousand globular clusters around the dwarf galaxy. Even though a dwarf galaxy would, in reality, contain only a few globular clusters \citep[see, e.g.,][]{Mateo1998,Peng2008} this does not affect the result as the globular clusters are treated as test-particles, therefore the globular clusters does not affect each other. Having so many globular clusters enables us to probe many different orbits for initial conditions for each simulation. In post-processing we then scale the fraction of globular clusters in different populations to a realistic number of globular clusters in a given dwarf galaxy

\subsection{Models for the galaxies}\label{sec:potential_models}
To model the potential field of M31 we use a modified version of the potential fitted to its rotation curve by \citet{Geehan2006}. Our M31 potential is a static, axisymmetric, bulge-disc-halo potential with a Hernquist model \citep{Hernquist1990} for the bulge, a Miyamoto-Nagai model \citep[][]{MiyamotoNagai1975} for the disc and a Navarro-Frenk-White (NFW) potential \citep[][]{NFW1996} for the halo. Compared to the model of \citeauthor{Geehan2006} we neglect the central black hole and fitted their exponential disc profile to the aforementioned Miyamoto-Nagai model. This simplification is reasonable since the encounters simulated for this work mainly orbit in the halo of the M31. As coordinates we used a Cartesian coordinate system in the simulations, using the conventional definitions for spherical radius $r^2 = x^2 + y^2 + z^2$ and $R^2 = x^2 + y^2$, with origin at the centre of M31 and $x,y$ in the plane of the galactic disc. For completeness the expressions for all models are given below in the order bulge-disc-halo:
\begin{equation}\label{eq:bulge_potential}
\Phi_{\rm b}(r) = - \frac{GM_{\rm b}}{r_{\rm b} + r},
\end{equation}
\begin{equation}\label{eq:MN_potential}
\Phi_{\rm d}(R,z) = - \frac{GM_{\rm d}}{\sqrt{R^2 + (a + \sqrt{z^2 + b^2})^2}}, 
\end{equation}
\begin{equation}\label{eq:NFW_potential}
\Phi_{\rm h}(r) = -\frac{4\pi G \delta \rho_{\rm c} r_{\rm h}^3}{r}\ln\left(\frac{r + r_{\rm h}}{r_{\rm h}}\right),
\end{equation}
where $G$ is the gravitational constant and the parameters were set to $M_{\rm b} = 3.3\times10^{10}\,{\rm M}_{\odot}$, $r_{\rm b} = 0.61\,{\rm kpc}$. $M_{\rm d} = 1.03\times10^{11}\,{\rm M}_{\odot}$, $a = 6.43\,{\rm kpc}$, $b=0.265\,{\rm kpc}$, $\delta = 27\times10^4$, $r_{\rm h} = 8.18\,{\rm kpc}$ and $\rho_{\rm c} = 277\,h^2\,{\rm M}_{\odot}\,{\rm kpc}^{-3}$. For the NFW model we assumed a value $h = H_0 / (100\,{\rm km}/{\rm s}/{\rm Mpc}) = 0.71$. 

A caveat of a static potential field is that it does not produce any tidal decay for orbiting systems. Since our simulations run for a significant fraction of the age of the Universe this could cause misleading interpretations of our results. For our simulations it is mainly an issue for the globular clusters that are captured in the M31 potential after being tidally stripped away from the dwarf galaxy, since the orbit they obtain afterwards will determine whether they could be observed as MGC1.  \citet{BinneyTramaine2nd2008} derived a simplified expression for the timescale for orbital decay due tidal friction given by 
\begin{equation}\label{eq:tidal_decay_timescale}
t_{\rm fric} = \frac{1.17}{\ln{\Lambda}}\frac{\mathcal{M}(r)}{M}t_{\rm cross},
\end{equation}
where $\ln{\Lambda}$ is the Coulomb logarithm describing deflection of material due to the gravitational force of the orbiting system with mass $M$. $\mathcal{M}(r)$ is the mass enclosed within radius $r$ of the major galaxy, in this case M31. The crossing time, $t_{\rm cross} = r_{\rm i}/v_{\rm c}$ (where $r_{\rm i }$ is the initial orbital radius and $v_{\rm c}$ is the circular speed), is the time required for the orbiting system to travel one radian. Using a typical value for $\ln{\Lambda} = 0.6$, the timescale for orbital decay of an object like MGC1 (assuming $r_{\rm i} = 200\,{\rm kpc}$ and $M = 10^5\,{\rm M}_{\odot}$) in our M31 potential, given by
\begin{equation}\label{eq:tidal_decay_timescale_MGC1}
t_{\rm fric} = \frac{1.17}{\ln{\Lambda}}\sqrt{\frac{r\mathcal{M}(r)}{G}}\frac{r_{\rm i}}{M},
\end{equation}
far exceeds the age of the Universe.

This work is based on the assumption that MGC1 originates from a dwarf galaxy which at some point had an encounter with M31 during which the MGC1 was tidally stripped away and then left orbiting M31. To model the dwarf galaxy we used a Plummer model \citep[see, e.g.,][]{Plummer1911, BinneyTramaine2nd2008}. The potential field of such a model is expressed as
\begin{equation}\label{eq:plummer_potential}
\Phi_{\rm s}(r') = \frac{GM_{\rm s}}{\sqrt{r_{\rm s}^2 + r'^2}}.	
\end{equation}
where $M_{\rm s}$ is the total mass of the dwarf galaxy, $r_{\rm s}$ is a scale length and $r'$ is radius from the centre of the dwarf. The Plummer model gives the dwarf a cored profile, with a core radius that is approximately $0.64r_{\rm s}$. For this work we tested four different dwarf galaxy masses listed in Table~\ref{tab:dwarf_models}. These give typical masses and shapes for Local Group dwarf galaxies \citep[see, e.g.,][]{Mateo1998}.

For each encounter that is simulated, we use a sample of $1000$ globular clusters distributed in the dwarf galaxy. To cover as many possible initial positions as possible in the space surrounding the dwarf galaxy, we assume that the distribution is homogeneous in Cartesian space, i.e. ${\rm d}N/{\rm d}r\sim r^2$. We constrain the minimum orbital distance for a globular cluster in the dwarf such that the globular cluster is not disrupted prior to the encounter. The maximum orbital distance was selected such that the globular clusters are not tidally stripped before coming within $200\,{\rm kpc}$ of the M31. The radial limits are listed for each dwarf galaxy in Table~\ref{tab:dwarf_models}. The globular clusters are given radial orbits initially, however, since clusters are initiated when the dwarf is located at the apocentre of its trajectory, which gives clusters some time to evolve before reaching the pericentre where the vast majority is stripped away, their eccentricities will mix before the encounter takes place. 

\begin{table}
\centering
 \caption{Masses, scale-radius and the limits within which globular clusters where distributed initially.}
 \label{tab:dwarf_models}
 \begin{tabular}{lcccr}
  \hline
$M_{\rm s}$ 	& $r_{\rm s}$ &  $r_{\rm min}$	&  $r_{\rm max}$ & $M_{1/2}$\\
${\rm M}_{\odot}$	& ${\rm kpc}$ & $\rm kpc$	&	 $\rm kpc$	& ${\rm M}_{\odot}$\\
\hline 
$10^7$ & 0.2	&   $0.13$	&	 $3.9$	& $5\times10^6$\\
$10^8$ & 0.4	&   $0.38$	&	 $8.27$	& $5\times10^7$\\
$10^9$ & 0.6	&   $0.76$	&	 $17.7$	& $5\times10^8$\\
$10^{10}$  & 0.8 & $1.19$	&	 $37.94$ & $5\times10^9$\\
  \hline
 \end{tabular}
\end{table}

\subsection{Constraining encounters with the tidal radius}\label{sec:tidal_radius}
As was mentioned previously, the observed MGC1 appears void of tidal features such as tidal tails, association to tidal streams or nearby dwarf galaxies. We will use this to constrain the encounters which could lead to the donation of a MGC1-like cluster. Firstly, we do not allow any encounters that can cause visible damage to M31 to be the progenitors of MGC1, although we find that any encounters close enough to cause significant damage to the luminous structure of M31 would be tidally shredded to a large extent. 

To track whether the dwarf galaxy sustains significant damage to its luminous structure  during the encounter we use an analytic prescription to track its tidal radius, i.e., the radius outside which material starts to get stripped from the dwarf galaxy. \citet{Read_et_al_2006} derived a general analytic expression for the tidal radius. Their expression depends on both the orbital properties of the satellite galaxy, as well as the orbits of the object within the satellite. \citet{Read_et_al_2006} showed that for isothermal spherical density profiles, $\rho_{\rm g,s} = A_{\rm g,s}r^2$ (using the subscripts $\rm g$ for M31 and $\rm s$ for the dwarf galaxy), the tidal radius can be calculated by
\begin{equation}\label{eq:tidal_radius}
r_{\rm t} \simeq \frac{\sqrt{\beta A_{\rm s}/A_{\rm g}}\left(\sqrt{\alpha^2 + 1 + r^2/\beta} - \alpha\right)}{1 + \beta / r^2},
\end{equation}
where $\alpha$ is a factor that depends on whether orbits of objects in the satellite are prograde ($\alpha = 1$), radial ($\alpha = 0$) or retrograde ($\alpha = -1$). $\beta$ is used to simplify the expression and is given by 
\begin{equation}\label{eq:tidal_radius_Gamma}
\beta = 2\frac{r_{\rm p}^2r_{\rm a}^2}{r_{\rm p}^2 - r_{\rm a}^2}\ln\left({\frac{r_{\rm p}}{r_{\rm a}}}\right),
\end{equation}
where $r_{\rm p}$ is pericentre distance and $r_{\rm a}$ is apocentre distance of the satellite trajectory. Note that this formula only holds for bound dwarf galaxy trajectories. Throughout this work we will use the smallest tidal radius, that is with $\alpha = 1$ (prograde), to track whether the dwarf galaxy is tidally stripped of stars.

\section{Tidal stripping as mechanism to produce MGC1}\label{sec:results}
The main result of this work is to determine whether it is possible to tidally strip globular clusters away from dwarf galaxies and capture them onto wide orbits around M31. To test this we simulate encounters between M31 and dwarf galaxies, in which we trace 1000 globular cluster trajectories drawn from a distribution that represents possible initial orbits. This method relies on the fact that globular clusters are treated as test-particles, thus do not effect each other. For each simulated encounter we compare the fraction of the total clusters in three different populations, plus one sub-population: clusters that were stripped away from the dwarf and bound to M31 (Captured); clusters that remained bound in the dwarf galaxy throughout the entire simulation (Retained); clusters left unbound to either potential fields (Unbound). The sub-population consist of Captured clusters that end up on orbits which are different compared to the dwarf trajectory and that at some point after the encounter places the clusters at the orbital distance of MGC1 ($200\,{\rm kpc}$).

\subsection{Single encounter}\label{sec:single_encounter}

\begin{figure*}
	\includegraphics[width=2\columnwidth]{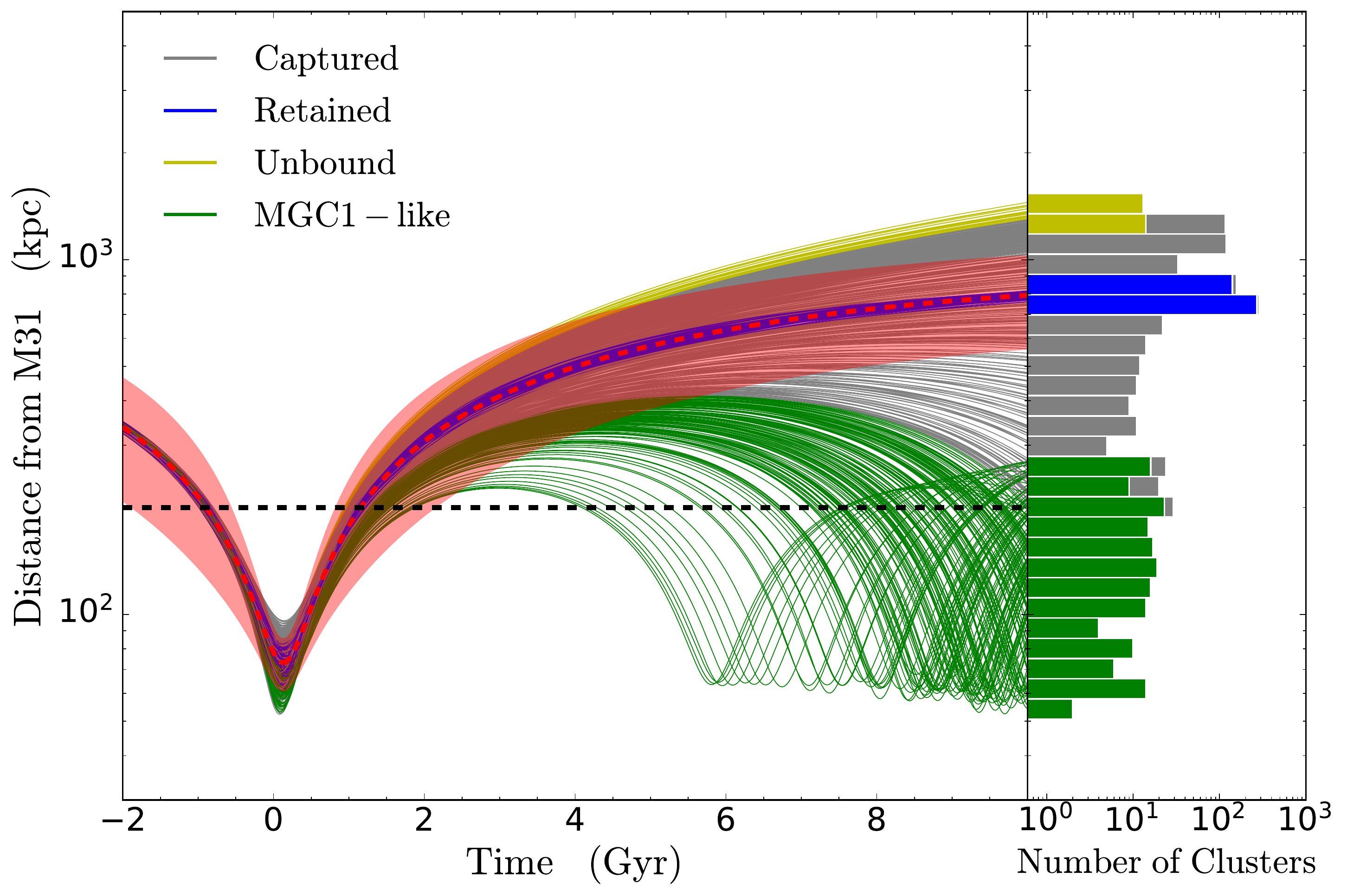}
    \caption{{\bf Left:} Distance from M31 as function of simulation time for the dwarf galaxy (red dashed line) and globular clusters (thin lines). Thin lines are coloured in accordance with the population in which the globular clusters end up in at the end of the simulation (see legend for labels). The distance between M31 and MGC1 is marked with the black dashed line. The red shaded region marks the tidal radius of the dwarf galaxy given by equation~\ref{eq:tidal_radius} assuming prograde orbits ($\alpha = 1$). {\bf Right:} Binned number of clusters at given distances from M31 at the end of the simulation ($t = 10\,{\rm Gyr}$). Bin height corresponds to the number of thin lines that hits the right-most boarder of the left plot. Bins are coloured similarly to colour of thin lines.}
    \label{fig:SVT_1e9}
\end{figure*}

In this section we describe a typical encounter in our simulations. We selected a $10^9\,{\rm M}_{\odot}$ dwarf galaxy with a trajectory that has a pericentre distance, $r_{\rm p} = 78.1\,{\rm kpc}$ and total specific energy, $E_{\rm tot} = -6.372\times10^3\,({\rm km/s})^2$ (corresponding to an apocentre distance of approximately 1 $\rm Mpc$). This encounter is chosen as one typical of those that produces MGC1-like clusters, see Fig.~\ref{fig:energy_dependence}. This encounter also showed the possible outcomes of different globular clusters in an encounter. As in all other encounter we trace the trajectories of 1000 individual globular clusters, which are initiated in a well mixed distribution before the encounter approached the smallest separation in its trajectory. At the end of the simulation we can then compare the fraction of all clusters which end up in the aforementioned groups. In the left plot of Fig.~\ref{fig:SVT_1e9} we show the distance from the M31 centre for the dwarf galaxy (red dashed line) as well as all the globular clusters (thin lines) as function of time for a typical encounter in our simulations. The thin lines are coloured depending on whether a given cluster ended up as Captured (grey), Retained (blue), Unbound (yellow) or MGC1-like (green) at the end of the simulation (in this case $10\,{\rm Gyr}$). The red region surrounding the dashed red line shows the extent of the tidal radius given by equation~(\ref{eq:tidal_radius}), and for this particular encounter we conclude that the encounter does not tidally disrupt the dwarf galaxy or cause visible damage to the M31. To the right in this figure we show a histogram for number of clusters at certain radii at the final time of the left plot (colours of the bars indicate the same as for thin lines).

\begin{table}
\centering
 \caption{Fraction of clusters in each group for a typical encounter. Data from same encounter as in Fig.~\ref{fig:SVT_1e9}}
 \label{tab:single_encounter}
 \begin{tabular}{lccr}
  \hline
Captured	&  Retained	& 	Unbound & MGC1-like \\
$0.561$	&	 $0.412$	&	 $0.27$	&	$0.206$ \\
\hline
  \hline
 \end{tabular}
\end{table}

For our set-up, the majority of the clusters are stripped away and captured, however, only a subset end up on orbits that makes them MGC1-like. Table~\ref{tab:single_encounter} displays the fraction of clusters in each group at the end of the simulation ($10\,{\rm Gyr}$ after pericentre passage) for the single encounter shown in Fig.~\ref{fig:SVT_1e9}. For this particular encounter $20\%$ of the globular clusters initially bound to the dwarf are tidally stripped away from the dwarf and end up on orbits that at some point places them $200\,{\rm kpc}$ away from M31. Note that for a cluster to be MGC1-like it has to pass an orbital distance of $200\,{\rm kpc}$ without being associated with the dwarf galaxy. For example, in Fig.~\ref{fig:SVT_1e9}, clusters passing $200\,{\rm kpc}$ the first time after the pericentre passage are not counted as being MGC1-like until a second passage as they would be associated with the dwarf the first time.

\begin{figure*}
	\includegraphics[width=2\columnwidth]{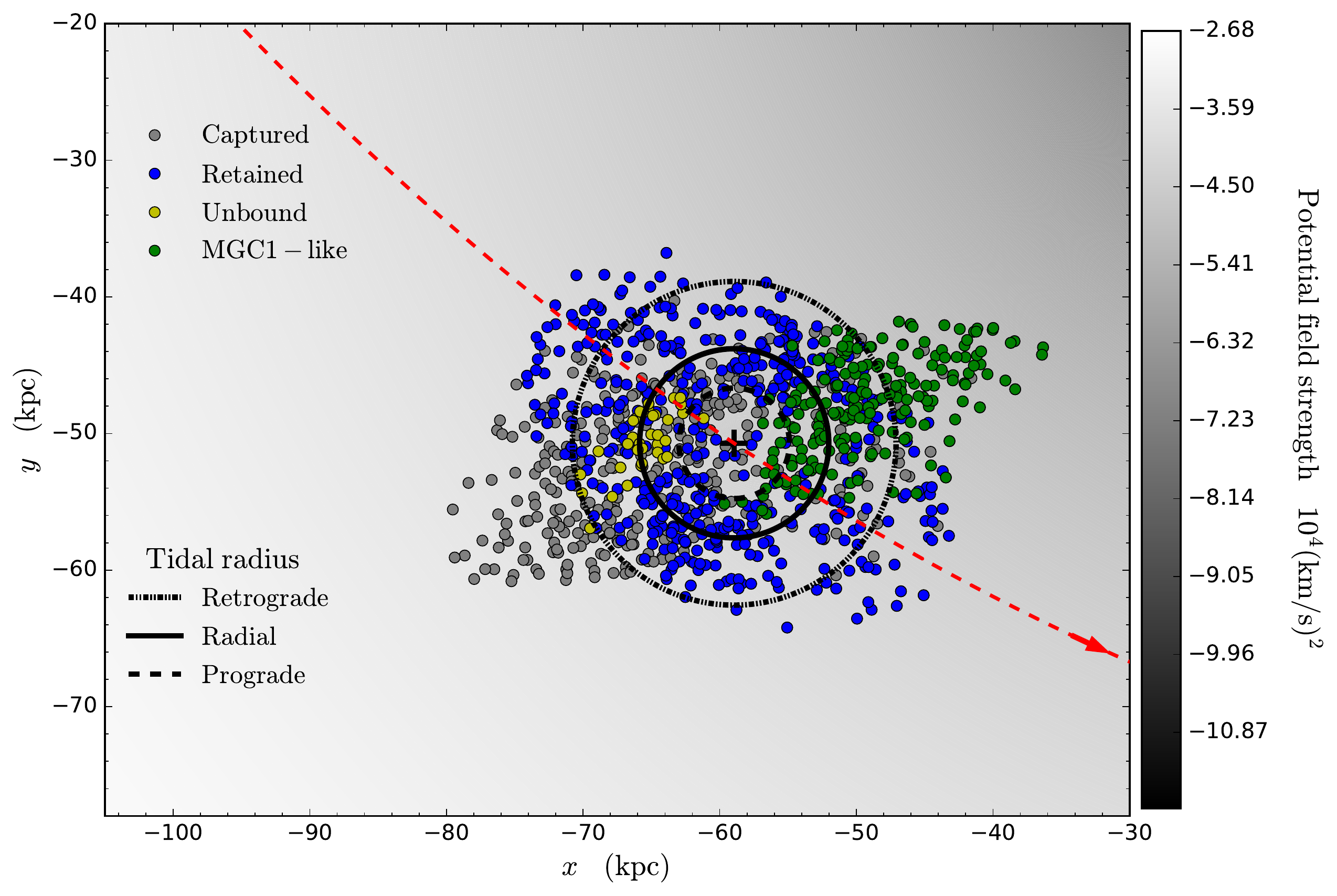}
    \caption{The spatial distribution of the globular clusters in the orbital plane of the dwarf galaxy when the dwarf galaxy is at its closest approach (i.e., $t=0$). The red dashed line shows the trajectory of the dwarf galaxy before (upper left) and after (bottom right) the encounter. The clusters are colour-coded according to what population they ended up in after the encounter (see legend for labels). The background shows the total potential field strength from the M31 galaxy and the dwarf galaxy. The black lines (dot-dashed, filled, dashed) shows circles with radius given by the tidal radius, given by equation~(\ref{eq:tidal_radius}), for retrograde, radial and prograde orbits.}
    \label{fig:cluster_pos_at_pericentre}
\end{figure*}

Is is interesting to see if certain properties are linked to whether a cluster end up in specific groups after the encounter, thus providing a means to predict the outcome. The tidal radius described earlier (see equation~(\ref{eq:tidal_radius})) indicate that prograde prograde clusters are more easily stripped away. Additionally, the position of a cluster when the dwarf galaxy passes the pericentre of its trajectory is also of interest as this will play a role in determining show their orbital energy is altered by the interplay between the dwarf and M31. Fig.~\ref{fig:cluster_pos_at_pericentre} shows the position of all clusters (points) as well as the dwarf galaxy (plus sign) at the pericentre passage. The trajectory of the dwarf is shown by the red dashed line. The grey-scale in the background shows the total potential field, that is the sum of the dwarf potential and the M31 potential. Note that the centre of M31 is located at (0,0) in this figure. The trajectory of the dwarf galaxy lies in the $xy$-plane. The first thing to notice is that the vast majority of the Retained clusters (blue) lie within the outermost tidal radius (retrograde orbit), as in indicated by the dot-dashed line. In fact, in the direction parallel to the trajectory of the dwarf they all fall within the distance equal to this tidal radius. Conversely, the majority of the clusters that were tidally stripped away during the encounter lie outside the innermost tidal radius (prograde). 

We find that the MGC1-like clusters mostly reside inside the orbit of the dwarf galaxy, whereas the Unbound clusters lie outside. The Unbound clusters lie rather deep into the dwarf potential as it passes the pericentre. The reason they are found close to the dwarf galaxy is because they are slingshot out from the system by the strength of its potential. The clusters outside the dwarf galaxy trajectory that are tidally stripped but left bound to M31 are those that end up on orbits exterior to the dwarf. It should be mentioned that if the dwarf galaxy is given a parabolic trajectory, then out of the clusters that are stripped away approximately half end up captured by M31 on orbits interior of the dwarf trajectory, while the other half are thrown away on hyperbolic trajectories exterior to the dwarf trajectory. 

\begin{figure}
	\includegraphics[width=\columnwidth]{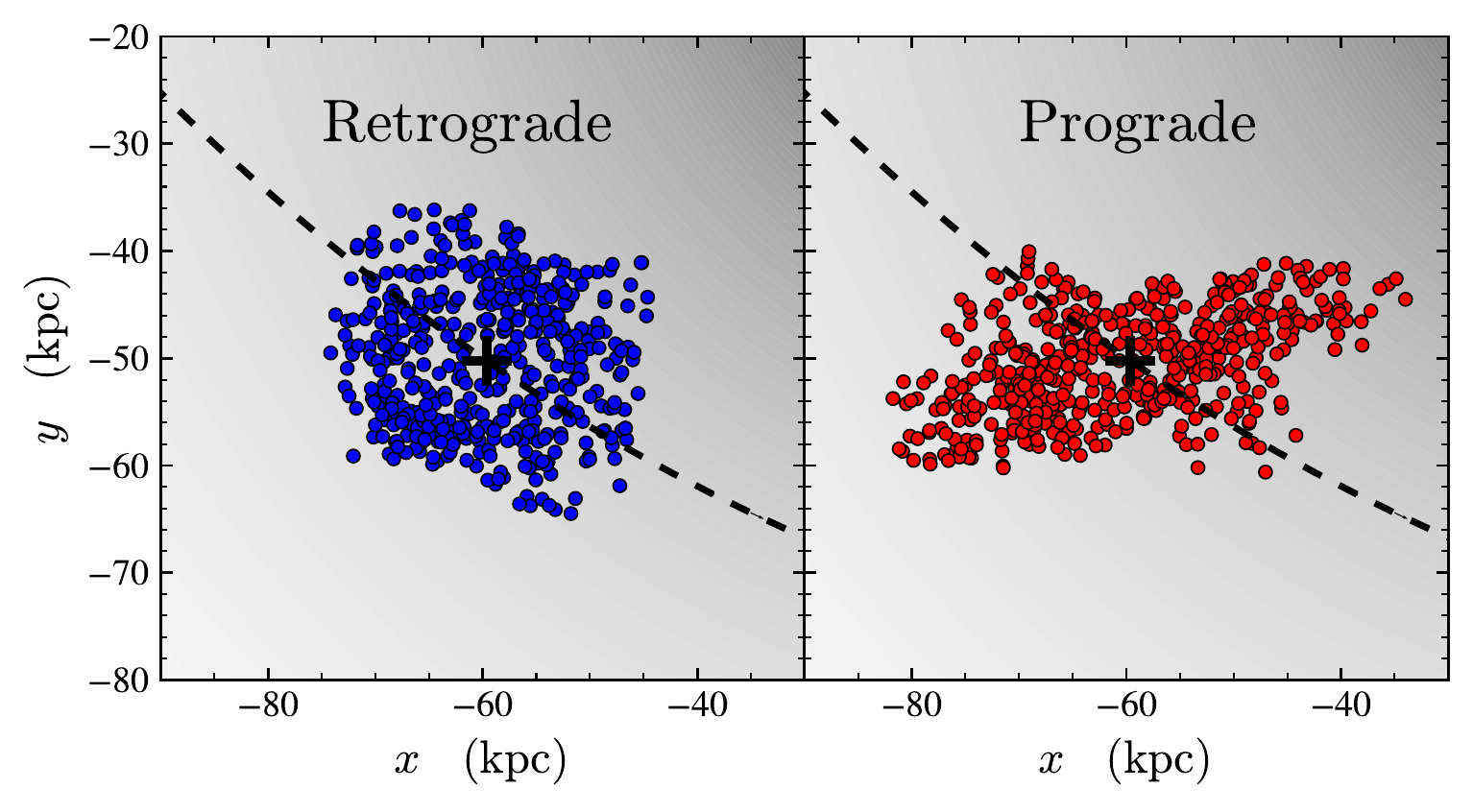}
    \caption{Plots similar to that in Fig.~\ref{fig:cluster_pos_at_pericentre}, but clusters are divided into clusters on retrograde orbits (left plot) and prograde orbits (right plot) when the dwarf is at the pericentre position.}
    \label{fig:retr_vs_prog_pos}
\end{figure}

Focusing now on the shape of the distribution, we find that the different cluster groups fall within overlapping but different regions. The Retained clusters are distributed in an elliptical shape with semi-major axis parallel to the trajectory of the dwarf, while the remaining clusters have a similar shape but more elongated and with semi-major axis oriented toward the direction of the M31 centre with a small offset angle. It turns out that one obtains the same shapes if dividing the clusters into those on retrograde orbits and prograde orbits, see Fig.~\ref{fig:retr_vs_prog_pos}. This clearly shows that tidal stripping not only depends on the tidal forces that is applied to certain objects, but also on their orbital properties as was shown by \citet{Read_et_al_2006}. 

\begin{figure}
	\includegraphics[width=\columnwidth]{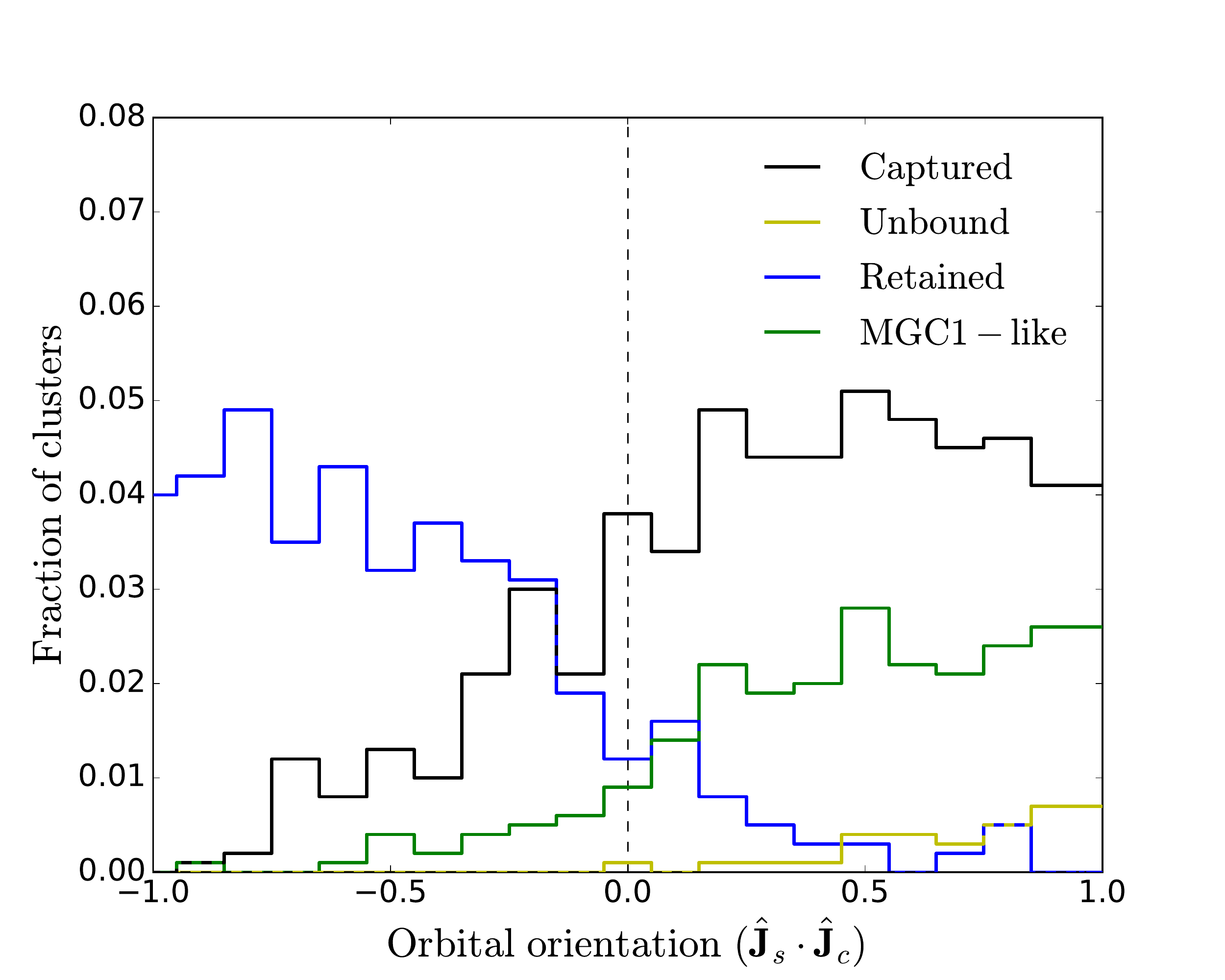}
    \caption{Fraction of clusters in different populations as function of the scalar product between normalised direction of the orbital plane for dwarf galaxy and globular cluster, see equation~(\ref{eq:cluster_orientation}). This is a measure of the degree to which a cluster has prograde (positive), retrograde (negative) or radial (zero) orbit with respect to the dwarf galaxy orbit around M31 at $t = 0$.}
    \label{fig:orbital_orientation_1e9}
\end{figure}

To investigate further we determined the degree to which each cluster is retrograde or prograde, quantified by the scalar product between the normalised angular momentum vector of the dwarf galaxy around M31 and that of each cluster orbiting around the dwarf given by the expression:
\begin{equation}\label{eq:cluster_orientation}
\hat{\mathbf{J}}_{\rm s} \cdot \hat{\mathbf{J}}_{\rm c} = \frac{\mathbf{r}_{\rm s}\times\mathbf{v}_{\rm s}}{|\mathbf{r}_{\rm s}\times\mathbf{v}_{\rm s}|} \cdot \frac{\mathbf{r}_{\rm c}\times\mathbf{v}_{\rm c}}{|\mathbf{r}_{\rm c}\times\mathbf{v}_{\rm c}|},
\end{equation}
where $\mathbf{r}_{\rm s}$ is the radial vector from M31 to the satellite, $\mathbf{v}_{\rm s}$ is the velocity vector for the satellite, $\mathbf{r}_{\rm c}$ is the radial vector from the satellite to the globular cluster and $\mathbf{v}_{\rm c}$ is the velocity vector of the cluster. This quantity will be positive for prograde orbits and negative for retrograde orbits. Fig.~\ref{fig:orbital_orientation_1e9} shows a histogram of fraction of clusters as function of orbital orientation, given by equation~(\ref{eq:cluster_orientation}), for the different cluster groups. This figure confirms that the clusters that are tidally stripped away are typically found on prograde orbits (positive orbital orientation), while retrograde clusters typically remain bound to the dwarf. Furthermore, we find that MGC1-like clusters are more likely among clusters that have an angular momentum vector which is closer to being aligned with that of the dwarf (orbital orientation closer to 1). This encounter has an apocentre which is significantly larger compared to the orbital distance of MGC1, thus some mechanism needs to bind them tightly to M31 otherwise they would leave the system along a trajectory similar to that of the dwarf. This mechanism is the interplay between the gravitational force exerted by the dwarf and M31. This can be thought of the opposite of the slingshot mechanism which throws the Unbound clusters out of the system.

\subsection{Likelihood of capturing MGC1}
To understand what kind of encounters are most likely to produce MGC1 we simulated a set of encounters covering a large range of dwarf galaxy trajectories. We tested four different dwarf galaxy masses, all listed in Table~\ref{tab:dwarf_models}. The initial conditions for the dwarf galaxy was set up by the following steps: (1) A pericentre position was randomly selected uniformly on a sphere with origin at the centre of M31 and a radius between $20\,{\rm kpc}$ and $120\,{\rm kpc}$, such that the distribution scales as $r_{\rm p}^2$. This is the same approach as was used when initialising the globular cluster sample, see Section~\ref{sec:methodology}; (2) The magnitude of the velocity of the dwarf was set by its specific kinetic energy, given by the potential energy subtracted from the total energy. The potential energy was readily available from the potential of M31 ($\Phi(r_{\rm p}) = \Phi_b(r_{\rm p})+\Phi_d(r_{\rm p})+\Phi_h(r_{\rm p})$), whereas the total specific energy was selected uniformly in the range [$\Phi(r_{\rm p}),E_{\rm max}$]; (3) To find the direction of the velocity we randomly selected the direction of the orbital angle, $\hat{J}$, (similarly to how the direction of $r_{\rm p}$ was generated) and used the cross product,
\begin{equation}
\hat{\mathbf{v}} = \frac{\mathbf{r}_{\rm p}}{|\mathbf{r}_{\rm p}|}\times\hat{\mathbf{J}}.
\label{eq:vel_direction}
\end{equation}
(4) When the position and velocity at the pericentre was selected the dwarf galaxy was integrated backwards in time out to the smallest of either the apocentre distance or $500\,{\rm kpc}$. At this point the globular cluster sample was added to the dwarf.
Motivated by test simulations we chose $E_{\rm max}=0$ for all encounters except those with a dwarf of mass $10^{10}\,{\rm M_{\odot}}$, where $E_{\rm max}=10^4\,({\rm km/s})^2$ to cover the range where encounters produced MGC1-like clusters. The simulations where run until $10\,{\rm Gyr}$ after the dwarf passed the pericentre position. In total we simulated 1000 encounters for each dwarf galaxy mass, each probing 1000 different globular cluster orbits using test-particles.

\begin{figure*}
	\includegraphics[width=2\columnwidth]{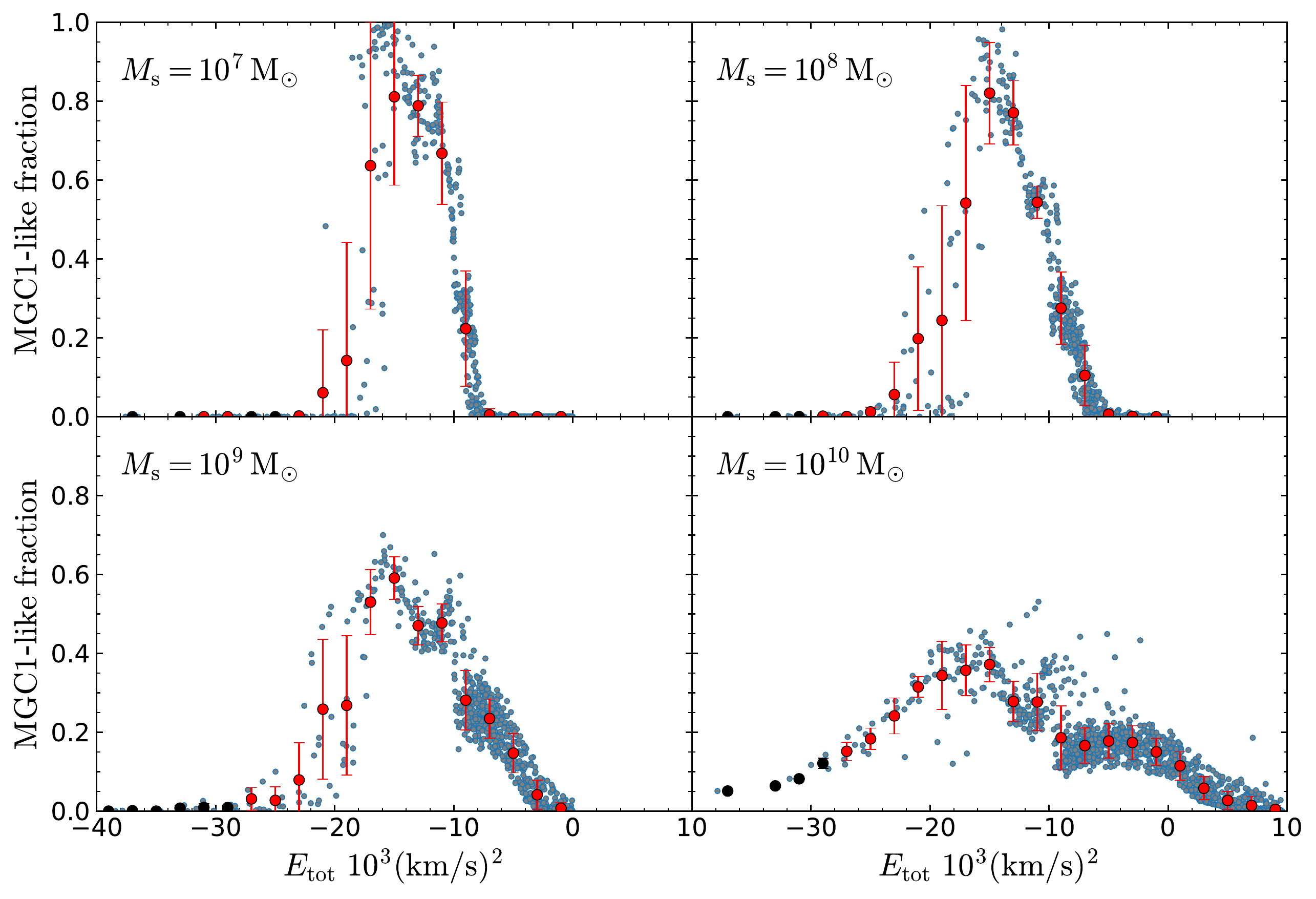}
    \caption{Fraction of MGC1-like clusters as function of specific orbital energy of the dwarf galaxy. Grey point show all individual encounters. The red points shows the mean fraction of all encounters in selected energy bins, with error bars that display the standard deviation. The black point show the same thing, however, in this energy bins where no more than 5 encounters where present. Each panel shows the results for different dwarf galaxy masses (indicated in each plot)}
    \label{fig:energy_dependence}
\end{figure*}

We find that there is a clear trend between fraction of MGC1-like clusters and the total specific energy ($E_{\rm tot}$), see Fig.~\ref{fig:energy_dependence}. This figure displays the fraction of MGC1-like clusters as a function of total specific energy for all simulations (grey points) for the four masses $10^7\,{\rm M}_{\odot}$ (top left), $10^8\,{\rm M}_{\odot}$ (top right), $10^9\,{\rm M}_{\odot}$ (lower left) and $10^{10}\,{\rm M}_{\odot}$ (lower right). The MGC1-like fraction peaks around $E_{\rm tot} \approx -15\times10^3\,({\rm km/s})^2$ then decreases toward both higher and lower energies at a rate that increases with decreasing dwarf galaxy mass. The peak is roughly at the orbital energy of MGC1, which is $\approx -19\times10^3\,({\rm km/s})^2$ assuming that MGC1 is on a circular orbit. Note that there are no known proper motions for MGC1, therefore an exact estimate of its orbital energy can not be determined. The red marks in Fig.~\ref{fig:energy_dependence} show the average MGC1-like fraction in different energy bins with widths of $2\times10^3\,({\rm km/s})^2$ centred on the red marks. The error bars show the standard deviation in the estimated average. The black points are the same as the red points, however, in these bins there were fewer than five data points. The standard deviation of the points are significantly smaller for specific energy bins larger that than the peak at $\approx -15\times10^3\,({\rm km/s})^2$, compared to lower specific energies. This is partly due to lower number statistics, but also due to how we set up the simulations. We do not consider any specific time when the dwarf is observed as a MGC1, but rather look at the possibility of being observed as MGC1. Since the simulations runs for $10\,{\rm Gyr}$ after the dwarf passes the pericentre of its trajectory, clusters that are stripped away and left on orbits with an apocentre only slightly larger than $200\,{\rm kpc}$ are almost certainly counted as MGC1-like. Additionally, dwarf galaxies on tighter enough orbits will have multiple encounters during which clusters can be tidally stripped away. 

The tidal radius described in Section~\ref{sec:tidal_radius} depends on distance between the dwarf galaxy and the M31 galaxy. Since the tidal radius is a good indication of whether clusters are stripped away or not it is interesting to look for relations between the MGC1-like fraction and the distance between the dwarf galaxy and M31, more specifically the pericentre distance since this will determine the tightest tidal radius. Assuming point mass potentials for both the satellite and the major galaxy, \citet{Read_et_al_2006} derived a simplified expression for the tidal radius at pericentre passage given by 
\begin{equation}
r_t \simeq r_{\rm p}\left(\frac{M_{\rm s}}{M_{\rm M31}}\right)^{1/3}\left(\frac{1}{1 + e}\right)^{1/3}\left(\frac{\sqrt{\alpha^2 + 1 + \frac{2}{1+e}}-\alpha}{1 + \frac{2}{1 + e}}\right)^{2/3},
\label{eq:pericentre_tidal_radius}
\end{equation}
where $r_{\rm p}$ is the pericentre distance and $e$ is the eccentricity of the satellite orbit (in our case the dwarf galaxy). This expression shows a similar scaling compared to the tidal radius derived by \citet{King1962}. This expression shows a linear relation between the pericentre distance and the tidal radius. Therefore, one would naively expect that the fraction of MGC1-like clusters depends on the pericentre distance of the dwarf trajectory. 

\begin{figure*}
	\includegraphics[width=2\columnwidth]{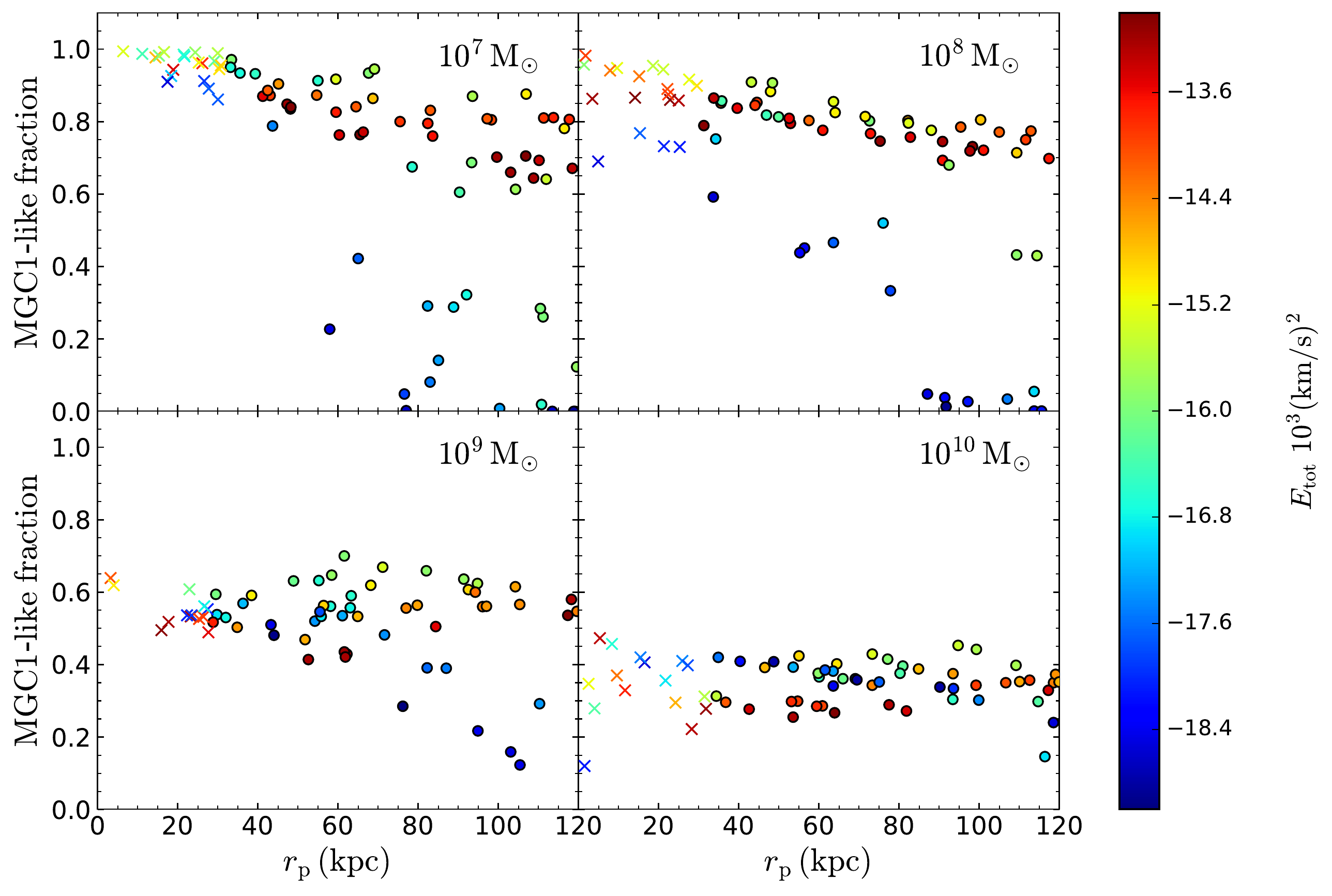}
    \caption{Fraction of MGC1-like clusters as function of pericentre position for the encounters with energies between $-20\times10^3\,({\rm km/s})^2$ and $-13\times10^3\,({\rm km/s})^2$, that is the energy range that covers the peak in MGC1-like fraction shown in Fig~\ref{fig:energy_dependence}. Points are coloured according to total specific energy in bins given by the colour bar. The four plots gives results from $10^7\,{\rm M}_{\odot}$ (top left), $10^8\,{\rm M}_{\odot}$ (top right), $10^9\,{\rm M}_{\odot}$ (bottom left) and $10^{10}\,{\rm M}_{\odot}$ (bottom right). The crosses marks encounters in which the dwarf galaxy sustained some damage during the encounter, whereas the dots marks encounters where the dwarf survive. Note that there is not a significant dependence between the MGC1-like fraction and pericentre distance, so long as the dwarf galaxy survives the encounter.}
    \label{fig:rp_dependence}
\end{figure*}

Fig.~\ref{fig:rp_dependence} shows the MGC1-like fraction as function of dwarf pericentre distance, $r_{\rm p}$, for the encounters with specific orbital between $-15\times10^3$ and $-9\times10^3\,({\rm km/s})^2$ for the four masses $10^7\,{\rm M}_{\odot}$ (top left), $10^8\,{\rm M}_{\odot}$ (top right), $10^9\,{\rm M}_{\odot}$ (lower left) and $10^{10}\,{\rm M}_{\odot}$ (lower right). We find a linear decrease in MGC1-like fraction with increasing dwarf galaxy pericentre in given energy ranges, although the slope is very small with the exception of small dwarf galaxy mass with bound orbits. For the larger masses we do not find any clear dependence with pericentre distance for this particular energy range. If looking at other energy bins, especially toward larger specific energy bins there is a dependence that is similar to the one found for the $10^7\,{\rm M}_{\odot}$ and $10^8\,{\rm M}_{\odot}$ dwarf galaxies, however, in these energy ranges the the fraction of MGC1-like cluster is negligibly small (a few per 1000 clusters) regardless of pericentre distance. 

In Fig~\ref{fig:rp_dependence} we separated encounters where the dwarf galaxy survive the encounter (i.e., has a luminous structure that is within the smallest pericentre distance, given by equation~(\ref{eq:pericentre_tidal_radius})) and encounters where the dwarf is tidally stripped of stars that do not belong to the cluster by filled circles and crosses respectively. For each mass there is a sharp cut in pericentre distance where this takes place due to the strong dependence between the tidal radius and pericentre distance (which set the smallest tidal radius). We find that to produce MGC1 from an encounter that does not leave behind any stellar stream due to tidally shredding the dwarf galaxy the encounter must have had a pericentre larger than $30\,{\rm kpc}$, $29\,{\rm kpc}$, $27\,{\rm kpc}$ and $31\,{\rm kpc}$ for dwarf galaxies of masses $10^7\,{\rm M}_{\odot}$, $10^8\,{\rm M}_{\odot}$, $10^9\,{\rm M}_{\odot}$ and $10^{10}\,{\rm M}_{\odot}$ respectively. That the limiting pericentre distance is similar for all masses comes from an interplay between massive galaxies retaining their stars more easily due to stronger gravitational field (thus have larger tidal radius) and larger luminous structure.

\section{Building the globular cluster population of M31}\label{sec:building_M31_GCpop}
\subsection{Accreted globular clusters in M31}\label{sec:accreted_clusters}
We find that for clusters captured by M31 there is no preferred orbital orientation, which implies that in the accreted population there should be the same number of clusters that have co-rotating orbits and clusters that have counter-rotating orbits with respect to the rotation of the M31 galaxy. In contrast the population that formed {\it in-situ} should have to some degree have retained the orbital orientation of the material from which it formed. This should give the {\it in-situ} population primarily co-rotating orbits. This dichotomy has been suggested as a means to separate the accreted and {\it in-situ} population before, see e.g. \citet{Brodie2006} for a review. For the purpose of this work we will use this to determine the number of accreted clusters in different regions of M31.

We make the following crude but simple assumptions: 1) all clusters that formed {\it in-situ} have retained orbits that are co-rotating with the rotation of the M31; 2) the accreted population consists of an equal number of co-rotating and counter-rotating clusters. We refer to number of co-rotating clusters as $\rm A$ and number of counter-rotating clusters as $\rm B$. The total number of clusters is the sum of $\rm A$ and $\rm B$. The total number of clusters can also be split into the sum of the total number of {\it in-situ} clusters and the total number of accreted clusters. From the first assumption it follows in a given group of clusters the number of {\it in-situ} clusters should be $\rm A - B$. From the second assumption it follows that in a given group of clusters the accreted number of clusters is two times $\rm B$. Following this we can calculate the number of accreted and {\it in-situ} cluster by the following expression:
\begin{equation}\label{eq:acc_vs_insitu}
{\rm A}+{\rm B} = \underbrace{{\rm A} - {\rm B}}_{\rm \it in-situ} + \underbrace{2\cdot{\rm B}}_{\rm accreted}.
\end{equation}
\begin{figure*}
	\includegraphics[width=2\columnwidth]{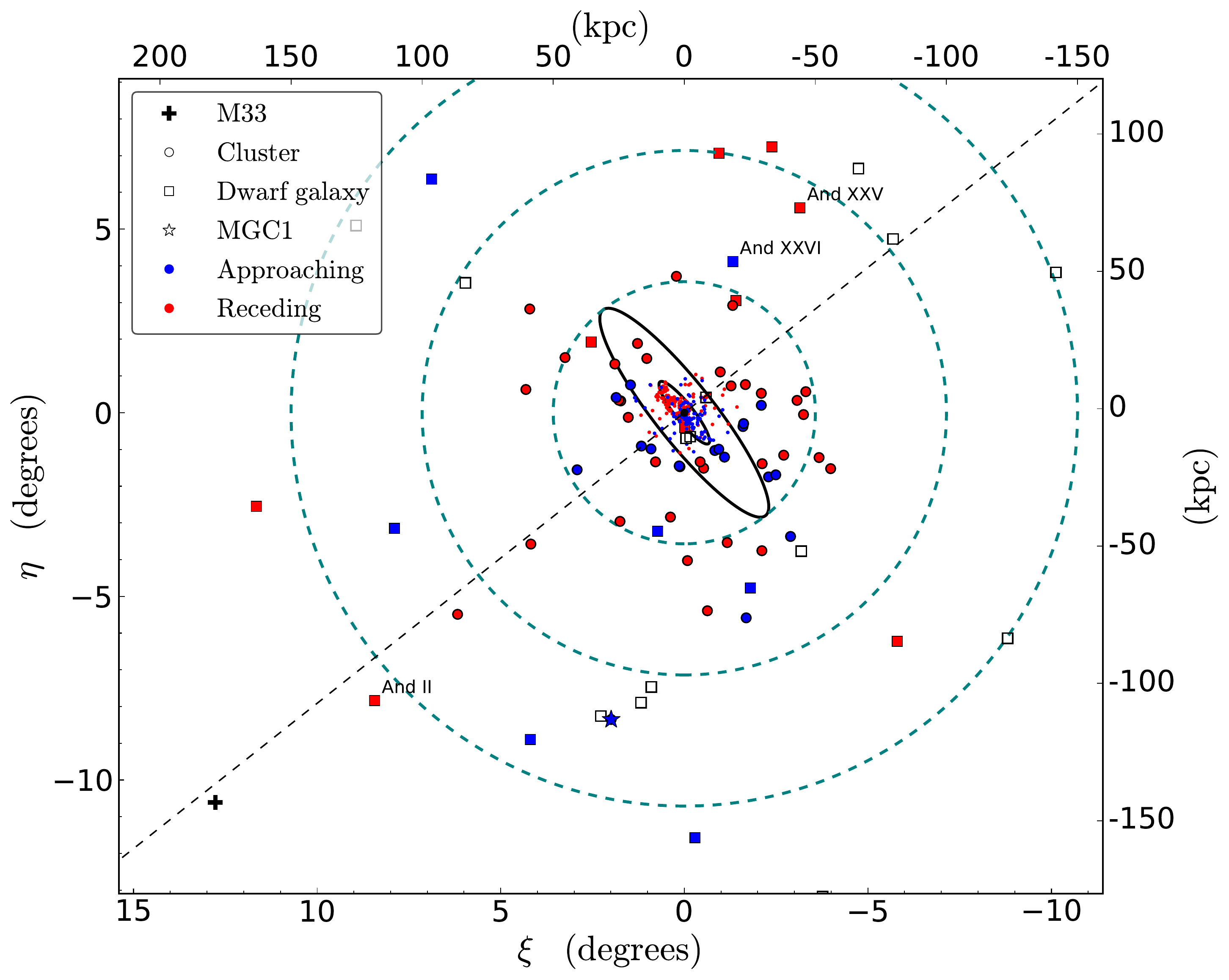}
    \caption{Position relative to M31 centre for observed globular clusters and dwarf galaxies. Objects are coloured depending on whether they are approaching (blue) or receding (red) with respect to the motion of M31. Clusters with unknown radial velocity are left unfilled. Cluster positions and velocities come from the M31 Revised Bologna Clusters and Candidates Catalog (Version 5), see \citet{Galleti2006} and \citet{Galleti2014}. Teal dashed lines show circles centred on M31 (black dot) with radii of $50\,{\rm kpc}$, $100\,{\rm kpc}$ and $150\,{\rm kpc}$. Black ellipses show projections of circles in the plane of the M31 disc with radii of $15\,{\rm kpc}$ and $50\,{\rm kpc}$. The black dashed line splits M31 along its polar axis. Note that the rotation axis of M31 points toward the bottom left of this plot. The satellite galaxies Andromeda II, Andromeda XXV and Andromeda XXVI are marked with names in the plot.}
    \label{fig:M31_GCpop}
\end{figure*}
In order to determine whether clusters in M31 are co-rotating or counter-rotating we looked at their position and radial velocity relative to the velocity of M31 to determine whether they are approaching or receding with respect to M31. Fig.~\ref{fig:M31_GCpop} shows the position of globular clusters and dwarf galaxies relative to the centre of M31. Clusters that are approaching relative to the M31 motion are coloured blue, while clusters that are receding are coloured red. Clusters or dwarf galaxies for which the radial velocity is not determined are left unfilled. All data that was used to create this figure was obtained from the M31 Revised Bologna Clusters and Candidates Catalog (Version 5), see \citet{Galleti2006} and \citet{Galleti2014}. In this frame of reference, M31 has a rotational axis that points in the direction of the dashed line toward the bottom left, i.e., the galaxy is rotating toward us on the lower right side of the dashed line and away from us on the opposite side. Following these conventions implies that clusters that are receding (red) on the upper left side of the dashed line and clusters that are approaching on the lower right side of the dashed line are co-rotating with the M31 rotation. The counter-rotating clusters will do the opposite. We divided the M31 into three different regions which are marked by black ellipses in Fig.~\ref{fig:M31_GCpop}. These ellipses are projections of circles that lie in the disc of M31. The innermost region is that with a projected circle with a radius of $15\,{\rm kpc}$ (Disc), the second region is that outside the first and second ellipse, where the second ellipse is the projection of a circle with a radius of $50\,{\rm kpc}$ (Inner halo). The last bin (outer halo) is the region outside the Inner halo. Using equation~(\ref{eq:acc_vs_insitu}) we determined the number of clusters in different groups for each of these regions and show the results in Table~\ref{tab:accreted_vs_insitu}. The errors that are shown is assumed Poisson noise in the determination of $\rm A$ and $\rm B$. 
\begin{table*}
\centering
\caption{Observed number of globular clusters in M31, divided into co-rotating (A) and counter-rotating (B) with respect to the M31 disc. Cluster were divided into different bins depending on their position according to ellipses given by the projection of circles in the plane of the M31 disc: BIN1 and BIN2 are projections of circles with radius $15\,{\rm kpc}$ and $50\,{\rm kpc}$, respectively, while BIN3 include the remaining clusters (see black ellipses in Fig.~\ref{fig:M31_GCpop}). Clusters where also divided into accreted and {\it in-situ} according to equation~(\ref{eq:acc_vs_insitu}).}
\label{tab:accreted_vs_insitu}
\begin{tabular}{lccccr}
\hline
Region	&	Co-rotating	& Counter-rotating	&	Accreted	&	{\it In-situ} & Fraction of {\it in-situ} \\
 &	A			&		B			&		$2\cdot{\rm B}$		& 		${\rm A}-{\rm B}$	 & $({\rm A}-{\rm B})/({\rm A}+{\rm B})$		\\
  \hline
Disc 	&	$141\pm 11.9$	&	$43\pm 6.6$		&   $86\pm 13.2$	&	$98\pm 13.6$	&	$0.533\pm 0.08$	\\
Inner halo	&	$51\pm 7.1$		&	$28\pm 5.3$		&	$56\pm 10.6$	&	$23\pm 8.9$		&	$0.291\pm 0.12$	\\
Outer halo	&	$22\pm 4.7$		&	$15\pm 3.9$		&	$30\pm 7.8$		&	$7\pm 6.1$		&	$0.189\pm 0.17$	\\
\hline
\end{tabular}
\end{table*}
In the last row of Table~\ref{tab:accreted_vs_insitu} we have listed the fraction of {\it in-situ} clusters for each region. We find that whilst roughly half of the innermost clusters have formed {\it in-situ}, the majority of clusters found further out have been accreted. That the number of {\it in-situ} clusters decrease significantly with increases distance is in agreement with other work, although we overestimate this fraction compared to estimates by \citet{Machey_Huxor2010} whom concluded that $\goa 80\%$  of clusters outside $15\,{\rm kpc}$ comes from accretion events, due to their spatial correlation with stellar streams. That we overestimate the fractions imply that the accreted clusters of M31 preferentially are on co-rotating orbits. This could be explained by satellite galaxies that have encounters with M31 preferentially come in on co-rotating orbits. In fact, if looking at the satellite galaxies that are shown in Fig.~\ref{fig:M31_GCpop} we find that there are more satellites on co-rotating orbits compared to counter-rotating orbits, although the numbers are very small. 

The accuracy errors that we estimate for the number of clusters in different regions should be discussed. The stated error are assumed Poisson noise which assumes that each event that contribute to the population should at least have similar properties, however, this is not necessarily the case. If M31 has had an encounter in the past that contributed with a large number of clusters then the orbital orientation of this encounters could gravely shift the assumption of an equal number of co-rotating and counter-rotating clusters in the accreted population. The reader should therefore note that these errors indirectly depend on the assumption that the accreted population has been built up by many encounters which has orbital directions that are isotropically distributed. 

\subsection{Constraining past encounters}\label{sec:constraing_past_enc}
Our numerical simulations shows that the MGC1 cluster most likely comes from an encounter where the dwarf galaxy had an orbital energy similar to that of the MGC1 today, given that the dwarf galaxy survived the encounter, regardless of the mass of the dwarf and its pericentre distance. Assuming that this can be generalised to encounters that produce globular clusters exterior to the orbital distance of MGC1, that is; for any encounter where globular clusters are tidally stripped away, the most likely outcome is having clusters on orbits with specific orbital energy close to that of the dwarf during the encounter. Assuming that MGC1 is the most distant cluster found in M31, this would imply that M31 is unlikely to have had encounters with dwarf galaxies that had a specific total energy larger than that of MGC1 ($\approx -19\times10^3\,{\rm km^2\,s^{-2}}$) during which globular clusters where added to the M31 cluster population. 

\begin{figure}
	\includegraphics[width=\columnwidth]{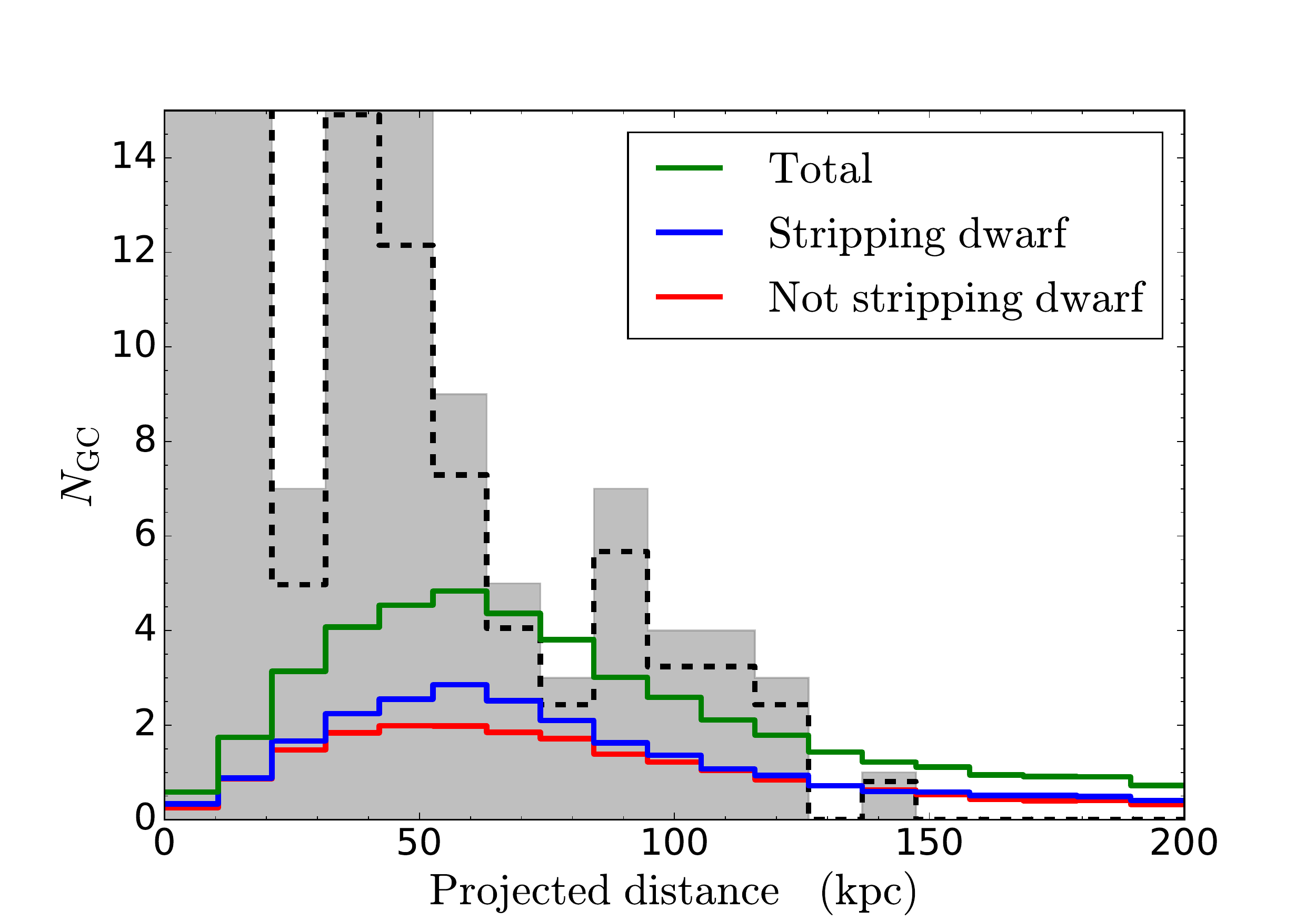}
    \caption{Projected number distribution of globular clusters around M31. Grey histogram shows the observed number distribution using data from \citet{CaldwellRomanowsky2016}, while the black dashed line shows number distribution of the fraction of the population that belongs to the accreted population based on the fractions displayed in Table~\ref{tab:accreted_vs_insitu}. Note that the innermost bins does not cover the full number within the number limits. The green line shows the distribution of clusters that was obtained by selecting encounters from our simulations until one cluster was obtain in the range $180-220\,{\rm kpc}$ (see details in Section~\ref{sec:constraing_past_enc}). The blue (red) line shows clusters originating from dwarfs that were (not) tidally stripped.}
    \label{fig:M31GC_number_distribution}
\end{figure}

It is interesting to compare the total contribution to the globular clusters population of M31 that our encounters would have given that MGC1 originates from one of our encounters. For this purpose we built up a mock sample of the globular cluster population by randomly selecting clusters from encounter in our simulations until they had produced 1 cluster at a distance from the M31 centre in the range $180-220\,{\rm kpc}$. The encounters were selected from all simulations in which the specific orbital energy of the dwarf was less than $-15\times10^3\,{\rm km^2\,s^{-2}}$. We found that selecting from encounters with energies larger than this produced a significant population of clusters outside $200\,{\rm kpc}$, which is in agreement with what was discussed previously. The encounters where selected such that the dwarf galaxy stellar masses matched the cumulative stellar mass function of satellite galaxies in the Eris simulation, see \citet{Pillepich2015}. To convert from our masses to stellar masses for the dwarf galaxies we used work by \citet{Ferrero2012}. For each selected encounter we selected a number of globular clusters from the entire initial population. Cluster that belonged to the Captured group where then added to the mock population. This selection was performed 1000 times and the average distribution of clusters is plotted in Fig.~\ref{fig:M31GC_number_distribution} with a green line. The number of clusters that was selected from each dwarf were drawn from different ranges depending on its  mass: $[1,3]$ ($10^7\,{\rm M}_{\odot}$), $[1,6]$ ($10^8\,{\rm M}_{\odot}$), $[1,10]$ ($10^9\,{\rm M}_{\odot}$) and $[1,25]$ ($10^{10}\,{\rm M}_{\odot}$) at a time after the encounter uniformly selected between $1-10\,{\rm Gyr}$. These ranges where selected based on the number of globular clusters that is observed in Local Group dwarf galaxies \citep{Mateo1998}. 

We now focus on comparing the spatial distribution of clusters that we obtain in the our simulations to that of the observed M31. We found that in order to obtain one cluster between $180-220\,{\rm kpc}$ in our simulation we need to select, on average, 10.3, encounters, however, the standard deviation of this average is large ($\sim 8$). The mean number of clusters that the encounters contribute with is 51.7 clusters and their projected radial distribution is shown in Fig.~\ref{fig:M31GC_number_distribution}. The grey area in the plot shows the projected spatial distribution of the observed M31 clusters, where the black dashed line marks the distribution of accreted clusters given by the method described in Section~\ref{sec:accreted_clusters}. The radial distribution of the observed clusters was obtained from \citet{CaldwellRomanowsky2016}. Note that MGC1 is located at a projected distance of $117\,{\rm kpc}$, resulting in a lack of clusters outside a projected distance of $150\,{\rm kpc}$. If the projected distance is interpreted as the radial distance, one would make biased conclusions. 

We find a surprisingly good match to the observed spatial distribution in the outer region, given that our method for building up the population is very basic. The simulated distribution peaks at around $50\,{\rm kpc}$ and falls toward a value of 1 at $200\,{\rm kpc}$ by construction. We do not produce clusters on close in orbits, which is likely due to a lack of dynamical friction and friction with gas that is absent in our simulations. Additionally, the model for the dwarf galaxy is static, and thus would not dissolve, regardless of whether the dwarf galaxy sustains tidal damage. This means that our simulations does not treat mergers in any meaningful way. Furthermore, we divided the total accreted population into clusters from encounters where the dwarf suffered significant damage (Stripping dwarf) and clusters where the dwarf remained intact (Not stripping dwarf), which is shown with blue and red lines respectively in Fig.~\ref{fig:M31GC_number_distribution}. We defined significant damage as encounters in which the tidal radius was within the half-light radius of the dwarf galaxy. We find that in the inner regions ($>75\,{\rm kpc}$) the majority of clusters originate from encounters where the dwarf galaxy sustained damage, thus could be associated with a stellar stream, whereas outside this region there is approximately an equal number of clusters in both populations. \citet{Machey_Huxor2010} investigated 61 clusters from the PAndAS sample located more than $30\,{\rm kpc}$ from the M31 centre and found that 37 lie on major substructure. Furthermore, they found that another 13 clusters coincided with cluster over-densities, thus likely originate from the same progenitor. Our comparison between clusters originating from stripped dwarf galaxies and isolated globular clusters does not account for over-densities between clusters, nevertheless, the fraction of clusters originating from encounters with tidally stripped dwarfs agree surprisingly well. The total number of clusters shown in Fig.~\ref{fig:M31GC_number_distribution} is $\sim 44$, out of which $\sim 24$ comes from stripped dwarfs. It should also be noted that we do not cover any major merger events in our simulations and comparing our number of clusters to the total number of observed clusters implicitly assumes that we cover the entire range of encounters that M31 has had in past. Since our encounters represent only a subset of the total number of encounter M31 has had in past, the sample of cluster either only represents a sub-sample of the total cluster population or the dwarf galaxies not covered in our simulations do not contribute with clusters to the M31 population. Furthermore, the total number of encounters (including mergers) one would expect for a spiral galaxy like the M31 far exceeds 10 \citep[see, e.g.,][]{Renaud2017}.

\section{Discussion}\label{sec:discussion}
In this work we investigate the possibly of tidally stripping away a single globular cluster from a dwarf galaxy and placing it on a wide orbit, thus producing a viable mechanism for placing the cluster MGC1 where it is observed today. We have focused on the fraction of clusters that end up on orbits that enables them to appear as MGC1 assuming that the initial clusters are distributed uniformly in a region restricted by the fact that they could not have been stripped prior to the encounter. How this distribution compare to the true distribution is very difficult to assess since their distribution can vary significantly between different dwarf galaxies. Additionally, we find in our simulations that the clusters that remain bound in the dwarf after the encounter change their distribution, typically extending further outward. Dwarf galaxies that have subsequent encounters will therefore be stripped of more and more globular clusters, although the dwarf galaxy is static by definition in our simulations. Furthermore, we restrain from quoting an exact number of clusters that you would typically contribute for a given dwarf galaxy because the specific frequency of globular clusters varies significantly for low mass galaxies. \citet{Spilter2009} suggested the empirical relation between the total mass in the globular cluster population of a galaxy and its total halo mass as a method for estimating the number of clusters expected for a given galaxy \citep[see also][]{Blakeslee1997, Harris2017, Renaud2018}. The relation is very steep (mass in clusters decreases quickly with decreasing halo mass) thus one expect significantly more clusters in more massive galaxies. Following this relation one expects few to non globular clusters for $10^9\,{\rm M}_{\odot}$ and $10^{10}\,{\rm M}_{\odot}$ whereas the smaller dwarf galaxies that we test likely never hosts globular clusters. However, other works suggest that the specific frequency for globular clusters turns around and increases for the smaller dwarf masses, see e.g., \citet{Peng2008}. Note that \citeauthor{Peng2008} point out that there is a very large scatter for the smaller masses, thus a good estimate for specific frequency is difficult to estimate in the low mass range. A good example of this variety is the dwarf galaxies in the Local Group. Out of the all the Local Group dwarf galaxies ($>76$) only 12 have been detected to host globular clusters \citep{Grebel2016}. Furthermore, dwarf that shows a large deviation to the empirical relation from \citet{Spilter2009} is the Fornax dwarf galaxy located in the Local Group, which has a mass enclosed within $1\,{\rm kpc}$ that has been estimated to $1.25\times10^8\,{\rm M}_{\odot}$ \citep{Kowalczyk2018}, has a population of 6 globular clusters \citep{Mateo1998}. Comparable to our Plummer masses this in between $M_{\rm s} = 10^8-10^9\,{\rm M}_{\odot}$).

One of the constrains that we place on the encounters from which MGC1 was tidally stripped away is that they should not produce any stellar streams or association to the dwarf galaxy, since non of these features are associated with MGC1. Up to this point we have not mentioned the possibility of producing MGC1 in an encounter where the dwarf is disrupted and subsequently hidden through dispersion of its stars. \citet{MackeyFerguson2010} applied star counts to the region surrounding MGC1 in search for structure, however, could not find any indications of stellar streams. Moreover, \citeauthor{MackeyFerguson2010} found MGC1 to have stars at very extended radius, with members of the cluster found out to $450\,{\rm pc}$ (possibly as far out as $900\,{\rm pc}$), and point out that MGC1 likely spent a considerable time in isolation. A next step in investigating the origin of the MGC1 would be to test its evolution using proper N-body treatment of its stars in the tidal fields that one expects from the encounters that typically produce clusters like the MGC1. Cluster evolution has been tested both for cases where the clusters is located in isolation \citep[see, e.g.,][]{Spitzer1987, Baumgardt2002}, as well as in tidal fields expected in accretion events \citep{Madrid2012,Miholics2016}, although, only in specific cases. Further investigating how the structure of clusters like the MGC1 evolve could provide useful constraints on otherwise invisible encounters between dwarf galaxies and larger galaxies. It should also be mentioned that if MGC1 did originate from an encounter where the dwarf was tidally shredded, low mass dwarf galaxies would be preferential, since they are more easily dispersed.

\subsection{Static potential fields}\label{subsec:disc_static_pot}
One of the shortcomings of our work is the simplicity of the potentials that were used. For the M31 we used a static potential, which was fitted to the rotating curve of the present-day M31. We have restricted our dwarfs to pass the pericentre of their trajectory at maximum $10\,{\rm Gyr}$ ago. At this time most of M31 should have been built-up if it had a evolution comparable to the Milky Way, \citet{Renaud2017,Kruijssen2018}. Concerning the dwarf galaxy one can question whether a static Plummer potential is realistic, since the dwarf will experience strong tides during the encounter which can cause significant deformation to its mass distribution and therefore its potential field. The goal of this work was to test a large set of encounters to determine a first estimate of the likelihood of producing extremely isolated globular clusters given certain orbital parameters. We tested four different dwarf galaxy masses for which 1000 different encounters, each testing 1000 different initial globular cluster orbits. Doing this in live dwarf galaxy potentials (i.e., treating the potential using tracer particles at some mass resolution) is beyond the scope of this work.

To investigate the effect of disrupting the dwarf we constructed a few simulations with a time-dependent dwarf potential which was gradually dispersed. We found that unless the dwarf is completely dispersed in extremely violent encounters, the orbits of the globular clusters that are stripped away does not change significantly. Typically, the encounters that do tidally disrupt the majority of the dwarf galaxy, have all clusters stripped away prior to this happening. As soon as they are stripped their trajectories are determined by the completely dominating potential of the M31. 

Furthermore, the dwarf galaxy is modelled using a Plummer potential. This choice was made since the Plummer model is a well behaved analytically described potential. Additionally, the Plummer model gives a cored mass profile which is typical for dwarf galaxies \citep[see, e.g.,][]{Moore1994, Flores1994,Governato2012}. Contrary to our choice of potential model, dwarf galaxies are typically modelled using a NFW profile, \citep{Navarro_etal1996}, although this model does not have a cored mass profile (a problem that can be alleviated using a cored-NFW profile \citep[see, e.g.,][]{Read_etal2016}). Another problem with the NFW profile is that the mass profile diverges for large radii. 

We investigate how our model compares to ones using a NFW profile by performing additional simulations with a dwarf modelled by the NFW profile, i.e., following the form already used for the M31 halo, see equation~(\ref{eq:NFW_potential}). For this comparison we used an orbit which is consistent with the known orbital properties of the M31 satellite ANDXXV. As we will see later, ANDXXV, which was discovered by \citet{Richardson2011}, is a good candidate as MGC1 progenitor. The coordinates used were taken from \citet{Galleti2006,Galleti2014}, which is shown in Fig.~\ref{fig:M31_GCpop} as a red square annotated with ANDXXV, at a projected position of $\xi\approx5.58$ and $\eta\approx-3.15$ degrees. Its heliocentric radial distance is $734\,\mathrm{kpc}$ and the heliocentric radial velocity is $107.8\,\mathrm{km}\,\mathrm{s}^{-1}$. Furthermore, ANDXXV has a mass comparable to those investigated in this work. An immediate problem with this choice is the lack of proper motion for this dwarf galaxy. For now we assume zero proper motion for ANDXXV since the focus is on comparing the models for the dwarf potenital. For both simulations we used the same initial radial distribution for the GCs in the dwarf, following the same method used in Sec.~\ref{sec:results}.  

We have to make additional assumptions to set the parameters for the ANDXXV models, since no estimate of its rotation curve exists in the literature. \citet{Kirby+2014} found the stellar mass within ANDXXVs half-light radius to be $5.4\times10^6\,{\rm M}_{\odot}$. This can be translated into a virial mass, $M_{200}$, in the range $10^9-10^{10}\,{\rm M}_{\odot}$ using empirical relations from abundance matching \citep{Read+2017,Read&Erkal2018}. However, such estimates are derived as a mean from a broad distribution, implying considerable uncertainty for individual galaxies. For our simulations we assumed a lower estimate of $M_{200}=10^9\,{\rm M}_{\odot}$ since this gives a model comparable to that in Sec.~\ref{sec:single_encounter}. Using a background density,  $\rho_c=136.05\,M_{\odot}\,\mathrm{kpc}^{-3}$ \citep{Maccio+2007}, and concentration parameter, $c = 22.5$, \citep[which is typical for dwarf galaxies, see e.g.,][]{Read_etal2016} we obtain a potential described by equation~(\ref{eq:NFW_potential}) using $\delta=3.34643\times10^5$ and $r_h=0.93\,\mathrm{kpc}$. A comparable Plummer potential was obtained by defining the mass to be the same at the virial radius, $r_{200}$, of the NFW halo. This gives a model according to equation~(\ref{eq:plummer_potential}) with $M_s=7.64\times10^8\,\mathrm{M}_{\odot}$ and $r_s=0.6\,\mathrm{kpc}$.

\begin{figure*}
  \centering
  \begin{tabular}{cc}
    \includegraphics[width=\columnwidth]{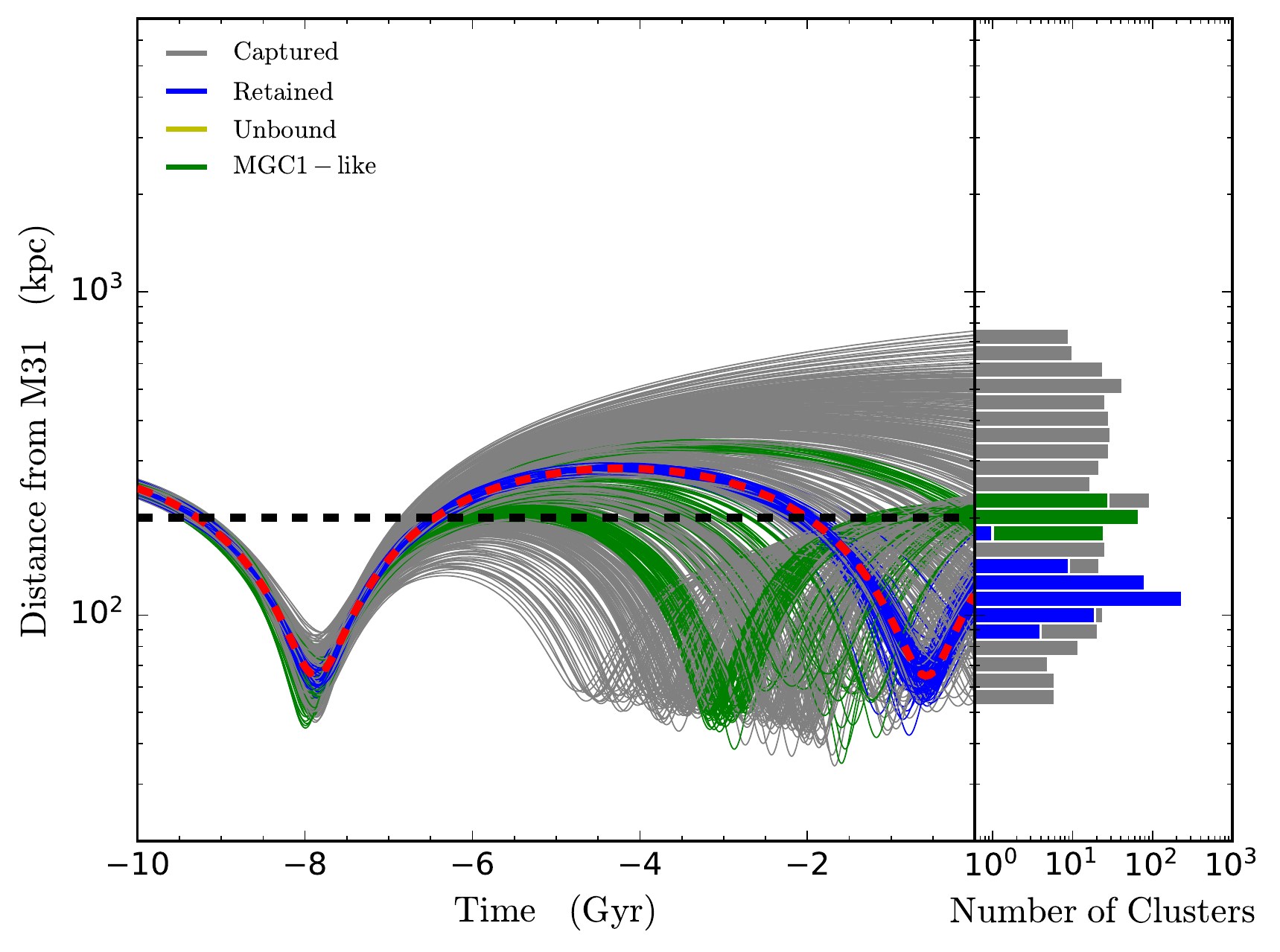} &
    \includegraphics[width=\columnwidth]{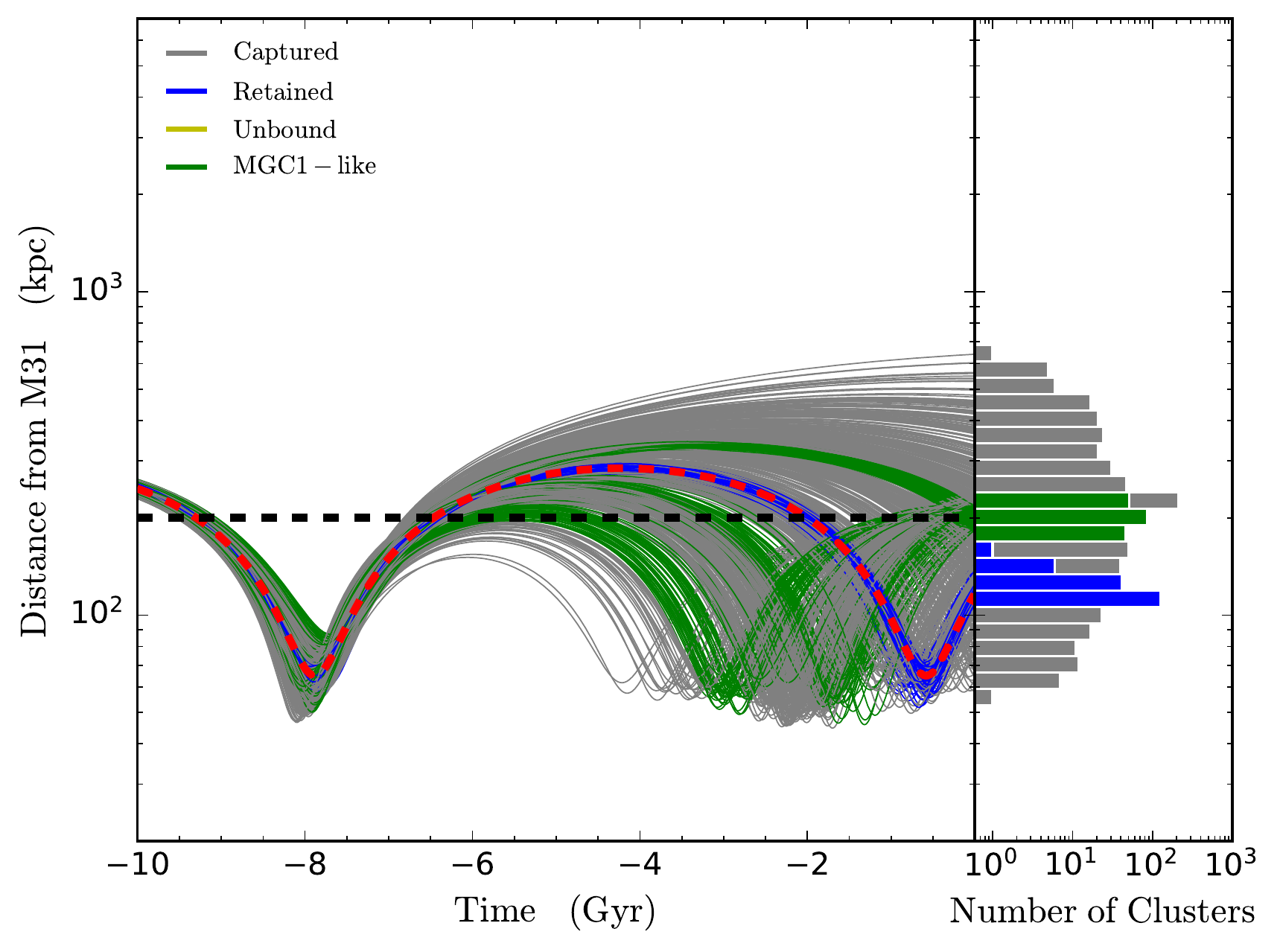}
  \end{tabular}
  \caption{Plots shows the same as Fig.~\ref{fig:SVT_1e9}, but for the trajectory of ANDXXV, assuming zero proper motion. Note that the time axis is set such that $t=0$ corresponds to the current location of ANDXXV. The left plot shows a simulation using a Plummer model for the dwarf potential, whereas the right plot is for one using a NFW model (see text for details).}
  \label{fig:SVT_model_comparison}
\end{figure*}

\begin{figure}
	\includegraphics[width=\columnwidth]{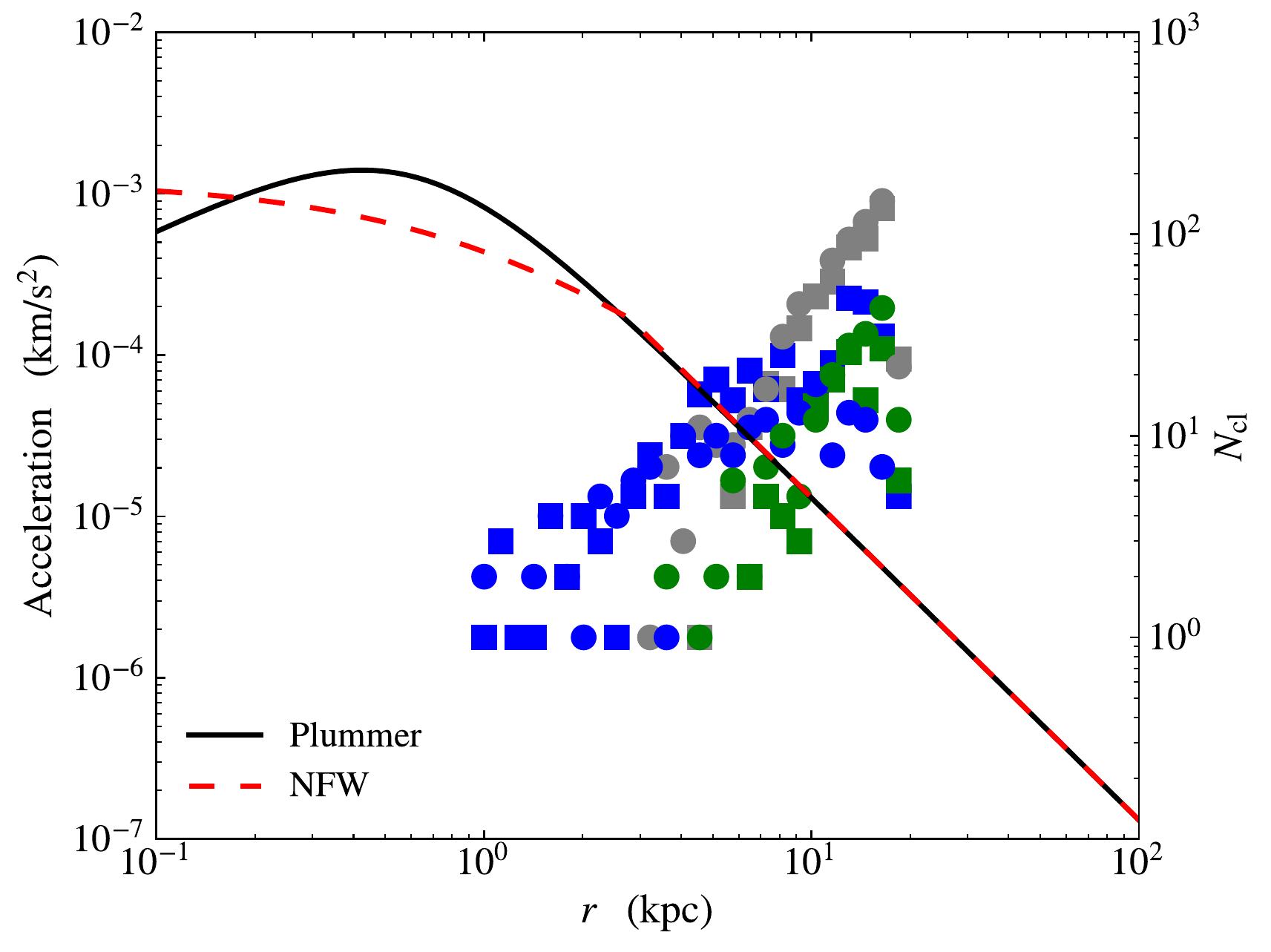}
    \caption{Figure comparing the acceleration (left axis) in the Plummer potential field with that of the NFW potential for the simulations an assumed ANDXXV orbit as function of radius, $r$, from the dwarf galaxy. The markers (squares for the Plummer simulation and circles for the NFW simulation) corresponds to number of clusters (right axis) at the initial position in the dwarf galaxy. Markers are colours according to where they ended up at the end of the simulation (colours are the same as for previous plots, see e.g, Fig.~\ref{fig:SVT_1e9}). }
    \label{fig:ic_model_comparison}
\end{figure}

Fig.~\ref{fig:SVT_model_comparison} shows trajectories and the final distribution of GCs of both the encounter using a Plummer model (left plot) and the one using the a model (right plot). The two simulation shows very similar distributions at the end, as well similar fractions of MGC1-like clusters ($12.1\%$ for Plummer and $18.2\%$ for NFW). There are however small differences in the shape of the wings surrounding the peak of the distribution. The GCs in the encounter with the Plummer model has a rather flat distribution with sharper edges, while the NFW model gives a GC distribution with a rounded shape. The reason for their similarity comes from the fact that the Plummer model and the NFW model is very similar in the region inside the dwarf galaxy where the GC are placed initially. This is shown in Fig.~\ref{fig:ic_model_comparison} where we plot the acceleration as function of radial distance from the dwarf galaxy for the two models. In this plot we included the initial radial number-distribution of GCs in the dwarf with markers coloured according whether they are Captured (grey), Retained (blue) or MGC1-like (green). We find that where the MGC1-like clusters are initiated the force from the two models are almost identical. It is mainly in the inner regions where the two models differ, where the Plummer has a cored density profile contrary to the NFW profile. Hence, when we simulate the same encounter with the two different models, the resulting GC distribution is the same.

\subsection{Possible progenitor of MGC1}
\begin{figure}
	\includegraphics[width=\columnwidth]{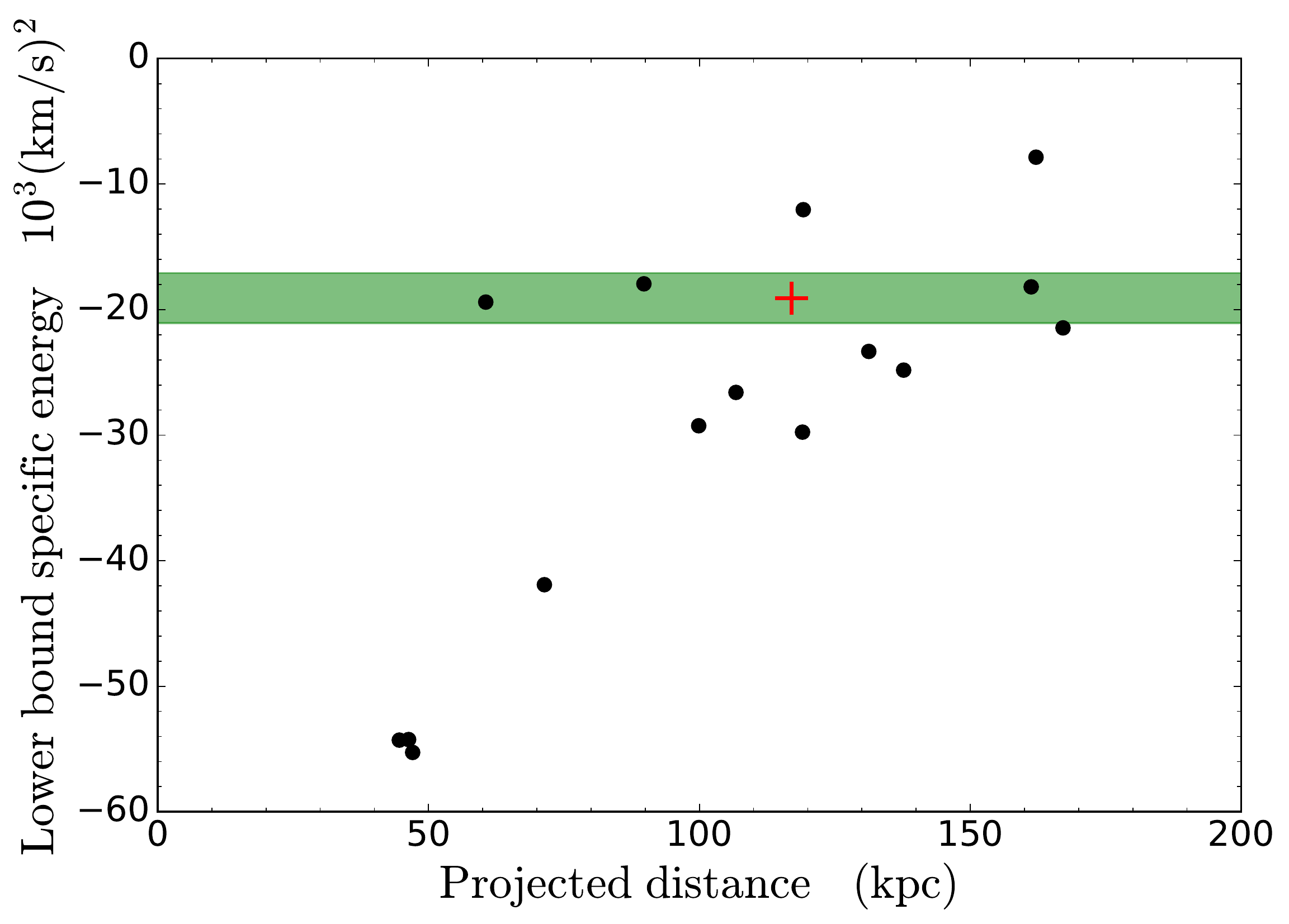}
    \caption{Lower bound specific energy as a function of projected distance for the observed satellite galaxies in M31. The energy is calculated by assuming that the radial velocity is the total velocity of the satellite and that the projected distance is the total distance between the M31 centre and the satellites. Because these quantities are only fractions of the total corresponding quantity the energy calculated must be a lower bound for the true specific energy. The plot also includes the same quantity for the MGC1 cluster (marked with red plus). The three satellites with lower bound specific energies closest to that of MGC1 (within the green region $\pm 2\times10^3\,{\rm km}^2\,{\rm s}^{-2}$ from the MGC1 marker) are Andromeda XXVII, Andromeda XXV and Andromeda II listed in order of increasing projected distance. These three satellites are named in Fig.~\ref{fig:M31_GCpop}, where we find that only Andromeda XXV has the same orbital orientation as MGC1. The satellite which is on the edge of the region is Andromeda XXIII.}
    \label{fig:satellite_energy}
\end{figure}

One of our main results is that if MGC1 comes from a dwarf galaxy, then the dwarf galaxy most likely had a specific orbital energy similar to that of MGC1. If the dwarf galaxy is still around then it must have remained on a fairly wide orbit, which means that any orbital decay due to dynamical friction would be minimal. This means that the dwarf galaxy probably retained its specific orbital energy. It is therefore interesting to investigate the satellite galaxies in M31 with this specific orbital energy. However, for the satellite galaxies in M31 we do not have 3D positions and velocities, but are restricted to radial velocities and projected distances. We can however estimate a lower bound energy, computed by using the projected distance and the model for the M31 potential used in our simulation to find the potential energy and the kinetic energy from the radial velocity alone. This energy is a lower bound because the projected distance must be less or the same as the total distance (therefore its potential energy will only increase as we add radial distance), and similarity the radial velocity is smaller or equal to the total velocity (therefore kinetic energy will increase as proper motion is added). In Fig~\ref{fig:satellite_energy} we plot this lower bound energy as a function of the projected distance for M31 satellites (black points) with known radial velocity and projected distance estimates. The plot also includes the values for the MGC1 (red plus) as well as green region which highlights the region $\pm 2\times10^3\,{\rm km}^2\,{\rm s}^{-2}$ from the MGC1 lower bound energy ($\approx -19\times10^3\,{\rm km^2\,s^{-2}}$). Within this region there are 3 satellite galaxies, Andromeda XXVII, Andromeda XXV and Andromeda II at the projected distances $60\,{\rm kpc}$, $90\,{\rm kpc}$ and $161\,{\rm kpc}$ respectively. These three galaxies are marked by name in Fig.~\ref{fig:M31_GCpop}, where we also find that only Andromeda XXV orbits on a co-rotating orbit. Because MGC1 is also on a co-rotating orbits, Andromeda XXV is a preferred choice for an MGC1 progenitor out of the three mentioned galaxies. This dwarf galaxy was investigated by \citet{Cusano2016}, whom found that Andromeda XXV consists of a single populations of stars with an age $9-13\,{\rm Gyr}$ and a low metallicity, $[{\rm Fe/H}]\sim-1.8$. Although, this is more metal-poor compared to the metallicity of MGC1 at $[{\rm Fe/H}]=-1.37\pm0.15$ \citep{Alves-Brito2009} they do have ages that are comparable.  
\begin{figure*}
  \centering
  \begin{tabular}{cc}
    \includegraphics[width=\columnwidth]{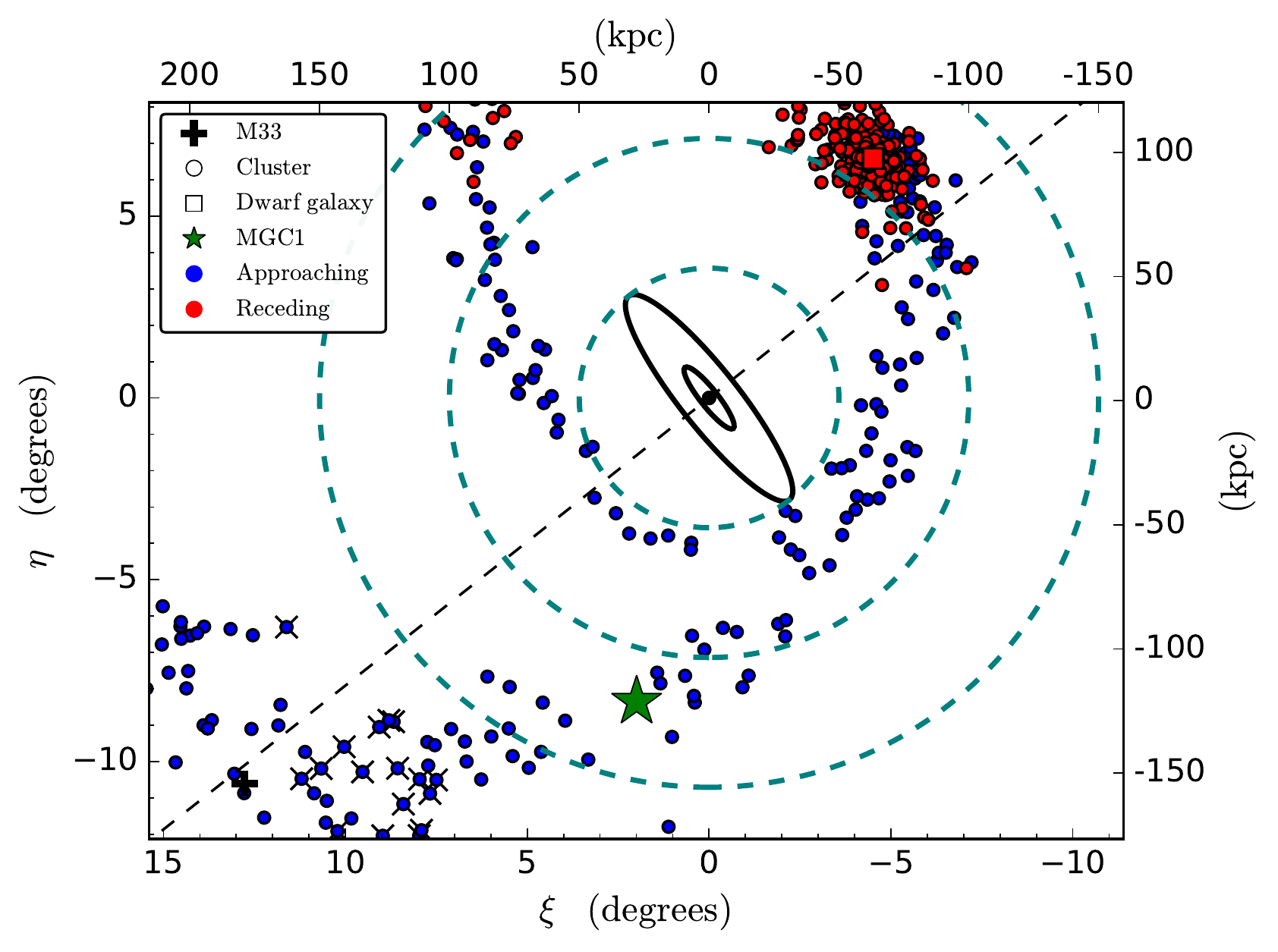} &
    \includegraphics[width=\columnwidth]{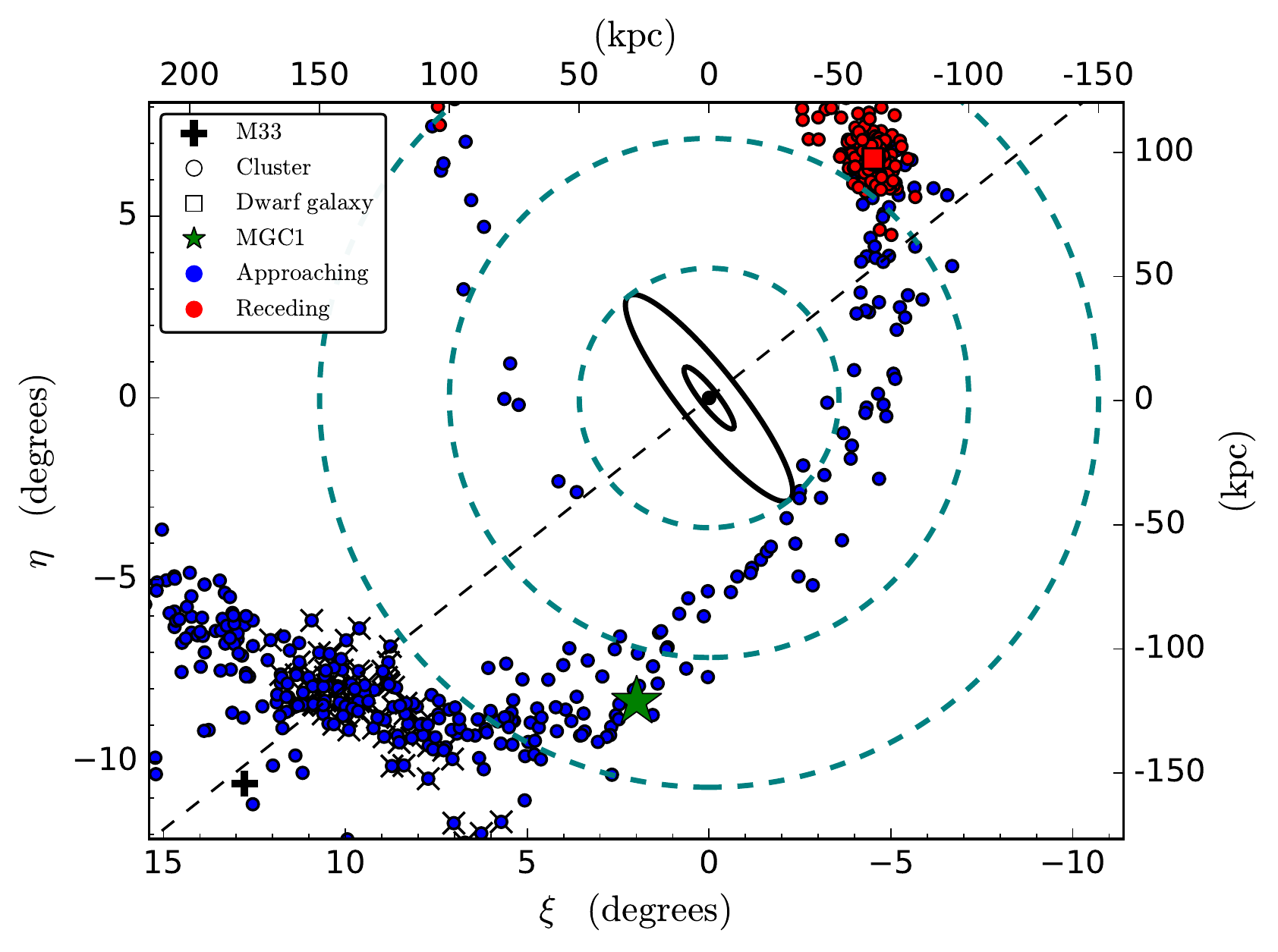}
  \end{tabular}
  \caption{Same as for Fig.~\ref{fig:M31_GCpop}, but for a simulated $10\,\mathrm{Gyr}$ evolution between orbits using the position and radial velocity of ANDXXV integrated backward $10\,\mathrm{Gyr}$ before starting for 1000 different GC trajectories initially bound to the dwarf galaxy. The dashed line divides the projected plane through the rotation axis (pointing toward the bottom left) of M31, which means that clusters co-rotate with M31 if approaching (blue) on the bottom right side of the dashed line or receding (red) its the top left side. The left plot shows the simulation using a Plummer model for the dwarf potential, and the right plot shows the simulation using a NFW model. Marked with crosses are clusters that are labelled as MGC1-like in Fig.~\ref{fig:SVT_model_comparison}. Note that the MGC1 is co-rotating with the rotation of M31, thus would appear blue in this figure.}
  \label{fig:andxxv_m31_projection}
\end{figure*}

Previously, when comparing the Plummer model with a NFW model, we chose to do so with an orbit that used the ANDXXV position and radial velocity as orbital parameters (assuming zero proper motion). In Fig.~\ref{fig:andxxv_m31_projection} we show the positions of all globular cluster trajectories after 10 Gyr evolution in the M31 potential (we integrated the dwarf 10 Gyr backwards before initialising the GCs). The figure includes both the simulation with the Plummer model (left) and the one with the NFW model (right). The distribution is comparable in both cases, although the GCs in originating from the NFW halo have a tendency to cluster in certain locations, whereas the GCs from the Plummer potential are more evenly spread out. The same is also seen in the histograms shown in Fig.~\ref{fig:SVT_model_comparison}. In Fig.~\ref{fig:andxxv_m31_projection}, clusters that are marked with blue filled circles on the bottom right of the dashed line co-rotate with M31, whereas the opposite is true for red filled circles. We also marked the GCs in this simulation that ended up on MGC1-like orbits with crosses. We find that very broadly these cluster end up in same part of the sky as MGC1. We stress however, that the without knowing the proper motion of either of their proper motions, we cannot say definitively that ANDXXV is the progenitor of MGC1 and this test shows only that this is possible. Better comparisons between metallicity, colour and similarities in stellar populations could provide a means for strengthening or excluding Andromeda XXV, as well as the other two mentioned dwarf galaxies as possible progenitors.

\subsection{Alternative origins for MGC1 and future work}
Finally, we want to focus on alternative origins of MGC1, which are not covered in this work. There has been previous work focusing on the formation of GCs in galactic encounters, see \citet{Renaud2018} and references therein. Several works has looked at formation and evolution of globular cluster in cosmological simulations, however, due to the high demand on computational power that is required to resolve the GC scale in such large simulation boxes, they are typically limited to the evolution of a single galaxy  \citep[see, e.g., ][]{Kravtsov&Gnedin2005,Renaud2017}. An exceptions is \citet{Hughes_etal2019} who looked at a suit of 15 zoom-in simulations of Milky Way-like galaxies from the E-MOSAICS project \citep{Pfeffer+2018}) with a focus on the properties of GCs and stellar streams from accretion of satellites. They suggest a method for estimating the mass of the progenitor satellite with a given GC given a reliable way of finding a stream associated with the GC. Unfortunately, with the case of MGC1, no such stream has yet been identified \citep{MackeyFerguson2010}. In our simulations we do not look at larger galaxies as progenitors, since these are more likely to leave streams. However, if MGC1 had a very wide orbit in its original host then it could have been stripped without much interaction between the two galaxies. One example of such an encounter would be the hypothesised encounter between M31 and M33 \citep{vanDerMarel2012, Brunthaler2005, Lockman2012}. M33 has a mass of $4.3\times10^{11}\,M_{\odot}$ \citep{Corbelli2014} and has been found to have a remarkable absence of an outer globular clusters population \citep{Cockcroft2011}. \citeauthor{Cockcroft2011} suggested that a possible explanation would be if M33 was heavily stripped of its globular clusters during an encounter with M31, which could have placed MGC1 where it is located today.

Another possible scenario for forming MGC1, which would not require any interaction with galaxies, is if it formed as an extra-galactic clusters which later accreted to M31. Extra-galactic cluster formation could occur in a cold filamentary stream from which gas was supplied to M31 in the very early Universe \citep{Mandelker2018}.

In this work we have focused on the globular clusters in M31, and in particular MGC1. However, since M31 has many similarities to Milky Way further work to compare our results to our own Galaxy could give useful results, although it would take this work beyond the scope of the main question addressed. In particular, the data available from the Gaia mission \citep[see, e.g.,][]{GaiaGC2018,Vasiliev2018} could give better understanding about how our method for distinguishing the number of accreted and {\it in-situ} clusters compare to other method, such as comparing blue and red clusters \citep{SearleZinn1978,MarinFranch2009,Keller2012}. Moreover, although M31 does not suffer from extinction in the same way as the Milky Way (especially in the outer parts), its globular clusters has the benefit of being significantly closer, and they could therefore be subject to very deep searches for association to tidal features. This could provide better means at determining the how rare extremely isolated clusters are.

\section{Summary \& Conclusions}\label{sec:conclusion}
We have tested if tidally stripping a globular cluster away from a dwarf galaxy during a close encounter with M31 and leaving it on a wide orbits without leaving any visible trace is a viable mechanism for producing the MGC1 cluster. We used simple models for the M31 galaxy and the dwarf galaxy in favour of running a large set of numerical simulations to test what type of orbital properties for the dwarf galaxy gives reasonably high likelihood of leaving behind clusters observable as MGC1. The main conclusions from our study are the following:
\begin{enumerate}
    \item We find that in certain ranges of specific orbital energy for a dwarf galaxy, there is a significant likelihood to produce isolated clusters on orbits wide enough to place them at an orbital distance comparable to that of MGC1. Regardless of mass the likelihood of producing such clusters peaks close to the specific orbital energy of MGC1 ($\approx-19\times10^3\,{\rm km}^2\,{\rm s}^{-2}$). Furthermore the amplitude of the peak varies for different masses (increasing with decreasing mass).
    \item In individual encounters, the clusters on prograde orbits are more likely to be tidally stripped away, in agreement with previous work for stars \citep{Read_et_al_2006}. To tidally strip away clusters and place them on orbits significantly different compared to the host dwarf galaxy (e.g. MGC1-like clusters) the angular momentum vector for the dwarf orbit around the major galaxy, $\mathbf{J}_{\rm s}$, and that of the cluster orbit around the dwarf, $\mathbf{J}_{\rm c}$, should align to as high degree as possible (quantified by $\mathbf{J}_{\rm s}\cdot\mathbf{J}_{\rm c}\approx 1$).
    \item We find that for the dwarf galaxies with smaller masses the likelihood of producing clusters like MGC1 decreases with increasing pericentre distance. Surprisingly, for more massive dwarf galaxies the decrease is small to negligible. Furthermore, we find that the dwarf galaxy trajectory needs a pericentre distance larger $\goa 30\,{\rm kpc}$ in order to survive the encounter regardless of mass. 
    \item In accordance with other works investigating accretion of globular cluster \citep[see, e.g.,][]{Brodie2006}, we find that there is not any preferred orbital orientation among the accreted clusters. Assuming that accretion of clusters leaves the same number of clusters on co-rotating and counter-rotating orbits, we are able to infer the relative frequency of accreted and {\it in-situ} clusters as a function of projected radius, see equation~(\ref{eq:acc_vs_insitu}) and Table~\ref{tab:accreted_vs_insitu}. For M31, we find in the inner region  roughly $50\%$ of the clusters have been accreted, whilst this figure increases to over $80\%$ further out. This agrees with the number found by \citet{Machey_Huxor2010}, whom counted accreted clusters by looking for association to stellar streams and spatial correlation between clusters.
\end{enumerate}
Furthermore, as a test that ties our simulations to observations of the global properties of M31, we compared the total number distribution of clusters that is accreted by M31 in our simulations to the distribution observed given that our simulations produce one cluster at the orbital distance of MGC1. We find that if we select encounters with specific orbital energy larger than the specific orbital energy of MGC1 then we tend to populate M31 with a significant number of clusters beyond $200\,{\rm kpc}$. Since MGC1 is one of the outermost cluster in M31 this implies that encounters between M31 and dwarf galaxies with specific orbital energy larger than this rarely contribute with globular clusters. Using this restriction for specific energy of the dwarf galaxy, but otherwise sampling encounters uniformly in specific energy with masses that match the subhalo mass function for a galaxy like M31, we find that the number distribution of clusters that are captured in our simulations matches the one observed in M31, with the exception of the innermost region where we do not contribute clusters. We do not populate the innermost region because our simulations do not cover encounters necessary to place clusters on such orbits. In addition, in some cases additional physics not covered in our model (for example dynamical friction) may play a role in shrinking globular cluster orbits.

\section*{Acknowledgements}

We thank Oscar Agertz and Louise Howes for contributing with useful comments in the early stages of this project. We thank the anonymous referee for helpful comments that improved our paper. We also acknowledge the Swedish National Infrastructure for Computing (SNIC) at Lunarc which provided resources and computer hardware for our simulations.




\bibliographystyle{mnras}
\bibliography{ref} 







\bsp	
\label{lastpage}
\end{document}